# Dual-polarized, mid-infrared nonreciprocal absorption

Simo Pajovic[1,2]*, Yiting Zhao[1], Yae-Chan Lim[1], Ruzan Sokhoyan[1], Harry A. Atwater[1]†

[1] *Department of Applied Physics and Materials Science, California Institute of Technology, United States of America*

[2] *Resnick Sustainability Institute, California Institute of Technology, United States of America*

*pajovics@caltech.edu

†haa@caltech.edu

**Abstract**

The emission and absorption of thermal radiation are usually coupled via reciprocity and Kirchhoff's law, stated as the equality of spectral directional emissivity and absorptivity. Magneto-optical materials have recently been identified as a promising route to lifting the constraint of reciprocity for thermal emission and absorption, with multiple experimental demonstrations using doped InAs, which is described by an antisymmetric, nonreciprocal dielectric tensor under an applied magnetic field. However, these demonstrations have been limited to *p*-polarized light in the Voigt configuration, whereas thermal radiation from a blackbody is unpolarized and thus comprised of both *s*- and *p*-polarized light. This means existing nonreciprocal emitter and absorber designs underutilize available power by 50%. Here, we report a photonic crystal slab-based structure that exhibits nonreciprocal absorption for both *s*- and *p*-polarizations in the mid-infrared spectral range. The structure operates from 11–20 μm and consists of an a-Si photonic crystal slab on top of a doped InAs substrate. The photonic crystal slab supports eigenmodes that couple to both *s*- and *p*-polarized plane waves, resulting in absorption peaks that, because of the applied magnetic field, spectrally shift in opposite directions for forward- and backward-propagating

light—a signature of nonreciprocity in planar, subwavelength systems. We fabricate the designed structure, then experimentally demonstrate dual-polarized nonreciprocal absorption by measuring its room-temperature absorptance using magnetic-field-integrated absorptance spectroscopy. Our design is a step toward the complete control of light as heat, which could improve photonic energy conversion, thermal management, and mid-infrared optical isolation and circulation.

## Introduction

Thermal emission originates from the random, microscopic fluctuations of the particles that constitute matter, pervading our daily lives and finding use in energy conversion, radiative cooling, and thermal camouflage. The application of nanophotonic design principles to thermal emission—also known as thermal photonics—has revolutionized our ability to mold the flow of light as heat[1]. Thermal photonics is enabled by Kirchhoff's law of radiation, which states that the spectral directional emissivity and spectral directional absorptivity of a surface are equal[2]. This equality facilitates the design and experimental characterization of thermal emitters, replacing time-consuming direct numerical simulations and noise-prone measurements of thermal emission with the much simpler problem of plane wave scattering and absorption[3,4]. Although Kirchhoff's law is useful, from a design point of view, it can be seen as a constraint because it couples thermal emission and absorption along identical spectral directional channels, making it so these quantities cannot be independently specified. In principle, overcoming Kirchhoff's law would open up a larger nanophotonic design space in which radiative heat transfer can be asymmetric and more degrees of freedom for optimization are available. Systems that are not constrained by Kirchhoff's law have potential applications in highly efficient solar energy harvesting[5,6] and thermal management, as well as opening up new avenues for nanophotonic design since emission and absorption can be decoupled.

One way to overcome Kirchhoff's law is to break Lorentz reciprocity by applying a magnetic field, which breaks time reversal symmetry and gives rise to antisymmetric components in the dielectric tensor of a gyrotropic material[7,8]. In these nonreciprocal systems, instead of equality between spectral directional emissivity $e_\ell(\omega, \theta, T)$ and spectral directional absorptivity $a_\ell(\omega, \theta, T)$, they are related by the difference in spectral directional reflectance $R_\ell(\omega, \theta, T)$ along opposite propagation directions:

$$e_\ell(\omega, \theta, T) - a_\ell(\omega, \theta, T) = R_\ell(\omega, \theta, T) - R_\ell(\omega, -\theta, T), \quad (1)$$

where $\ell$ is the polarization of light ($s$ or $p$), $\omega$ the angular frequency, $\theta$ the polar angle of incidence, and $T$ is temperature[9–11]. Equation (1) has now been experimentally validated numerous times[8,12–17], in particular using doped, n-type InAs, which is a popular and experimentally accessible gyrotropic material that supports free-carrier magneto-optical effects. Unstructured, bulk InAs can exhibit nonreciprocal thermal emission and absorption[9,15], but this nonreciprocity can be enhanced with the aid of nanophotonic resonances. Various modalities of nonreciprocal thermal emission and absorption have been experimentally demonstrated, including grating couplers based on guided-mode resonances (GMRs), which offer relatively strong, narrowband nonreciprocity[8,12], and multilayer, gradient epsilon-near-zero (ENZ) structures based on Berreman modes, which offer broadband nonreciprocity[13,14]. These GMR and gradient ENZ designs illustrate the potential of nanophotonics for enhancing and controlling nonreciprocal thermal emission and absorption due to applied magnetic fields.

A limitation of magnetic field-based approaches, however, is that the presence of nonreciprocity depends on the polarization of light, whereas thermal emission is intrinsically unpolarized. Prior work has focused on $p$-polarized light in the Voigt configuration (where the applied magnetic field $\boldsymbol{B}$ points perpendicular to the plane of incidence). The reason for this emphasis on $p$-polarized light is that the reflectance of $s$-polarized light is fundamentally reciprocal. Since its electric field is parallel to $\boldsymbol{B}$, the Lorentz force is not exerted on free carriers undergoing dipole oscillation in response to the electric field of incident light, as shown in Fig. 1(a). Thus, the optical response of $s$-polarized light is unaffected by $\boldsymbol{B}$. In principle, this means state-of-the-art designs can facilitate nonreciprocal control of only 50% of unpolarized, thermally emitted light. Therefore, expanding operation to both *p* and *s* polarizations is critically important for realizing the full potential of nonreciprocal thermal emission.

In the field of thermal photonics, it is well-known that structuring the surface of an emitter or absorber with a two-dimensional, periodic array of subwavelength scatterers allows for arbitrary polarization control in thermal emission[18–22]. In particular, cylindrical scatterers have been utilized to enable dual-polarized nonreciprocal thermal emission and absorption, but almost all prior work on this topic is theoretical and has postulated using InAs thin films on top of back reflectors[23–26] when, in practice, thin films of InAs can be difficult to grow on top of standard back reflector materials such as Al. Hence, there is a gap between theoretical predictions and realistic, fabricable InAs-based structures.

Here, we experimentally demonstrate dual-polarized, nonreciprocal absorption in the long-wavelength mid-infrared spectral range (11–20 μm) using an a-Si photonic crystal (PhC) slab on top of an InAs wafer. In our design, the PhC slab supports eigenmodes that couple to both *s*- and *p*-polarized light. The fringing fields of these eigenmodes penetrate the InAs, providing sufficient mode overlap with the anisotropic, magneto-optical medium to enable a nonreciprocal optical response. The applied magnetic field nonreciprocally shifts the ENZ frequency of the InAs, and as a result, the PhC slab eigenmodes become nonreciprocal as well. We fabricate our design using electron-beam lithography, then measure its room-temperature absorptivity using magnetic-field-integrated absorptance spectroscopy, experimentally demonstrating nonreciprocal absorption of both polarizations of light. Our result is a step toward the complete, nonreciprocal control of light as heat, which has the potential to enable more efficient energy conversion by decoupling emission and absorption along the same channels, better thermal management via directionally asymmetric thermal emission, and improved mid-infrared optical isolators and circulators. We note that during the preparation of this manuscript, we became aware of another experimental demonstration of dual-polarized nonreciprocal absorption, achieved at longer wavelengths (19-27 μm)[27].

## Results

*Physical mechanism of s-polarized nonreciprocity*

The principle of operation of our design is illustrated in Fig. 1. Our design operates in the Voigt configuration, in which the applied magnetic field $\boldsymbol{B} = B\hat{\boldsymbol{y}}$ points perpendicular to the plane of incidence ($\hat{\boldsymbol{y}}$ is the unit vector in the $y$-direction). This is convenient because (1) nonreciprocity for this direction of $\boldsymbol{B}$ is the strongest compared to all other possible directions because of maximum inversion symmetry breaking and (2) InAs happens to support $p$- and $s$-polarized bulk eigenmodes in the Voigt configuration, meaning there is no polarization conversion between $p$ and $s$ and analysis is greatly simplified (whereas in other configurations, the bulk eigenmodes are elliptically polarized, and analysis is considerably harder).

First, consider a planar, unstructured InAs surface, Fig. 1(a). The electric field of $s$-polarized light, $\boldsymbol{E}_s$, is parallel to $\boldsymbol{B}$ in the Voigt configuration. This means the dipolar oscillations of electrons inside the InAs, induced by $\boldsymbol{E}_s$, are also parallel to $\boldsymbol{B}$ and thus unperturbed (i.e., by the Lorentz force, which points in the direction of the cross-product of the electric field and $\boldsymbol{B}$), as shown on the left side of Fig. 1(a). In other words, $s$-polarized light does not "see" or interact with the applied magnetic field at all. By contrast, the electric field of $p$-polarized light, $\boldsymbol{E}_p$, is perpendicular to $\boldsymbol{B}$, and the induced dipole oscillations take the form of cyclotron motion, as shown in the right panel of Fig. 1(a). If we now introduce a periodic array of cylindrical voids in the $xy$-plane to break continuous translational symmetry, $s$-polarized light must satisfy the boundary conditions of Maxwell's equations—namely, continuity of the tangential components of the electric field—at the curved interface between air and InAs. This is shown in Fig. 1(b). This boundary condition requires that $\boldsymbol{E}_s$ gains components perpendicular to $\boldsymbol{B}$ as it transmits through the curved interface because its surface normal has both $x$- and $y$-components, thus enabling nonreciprocal interaction between $\boldsymbol{E}_s$ and $\boldsymbol{B}$. The periodic array of cylindrical voids can be designed to be a PhC slab, as illustrated in Fig. 1(c). We postulate that in the presence of a magnetic field and magneto-optical substrate such as InAs, this structure could support dual-polarized nonreciprocal scattering (i.e., weaker reflection in one direction and stronger reflection in the other).

Another way of understanding how our design works is illustrated in Fig. 1(d). The PhC slab has $C_2$ and $C_4$ symmetry, so the solutions of Maxwell's equations in the PhC slab take the form of sets of orthogonal Bloch eigenmodes that must also be $C_2$- and $C_4$-symmetric. We call these sets of eigenmodes TM (electric field parallel to the $xy$-plane) and TE (electric field normal to the $xy$-plane) eigenmodes. $C_4$ symmetry in particular allows the electric field of the TM eigenmodes to possess both $x$- and $y$-components, meaning the fringing, $xy$-polarized fields can interact with the applied magnetic field in the $y$-direction acting on the InAs substrate below the PhC slab. This is because the $x$-component of the TM eigenmode is perpendicular to the applied magnetic field. Therefore, the TM eigenmodes of the PhC slab become nonreciprocal, and they can be excited by $s$-polarized light via mode overlap since $\boldsymbol{E}_s$ has only a $y$-component. In turn, the absorptance of $s$-polarized light is nonreciprocal, manifested by the nonreciprocal TM eigenmodes.

Our understanding—based on the excitation of the TM eigenmodes of the PhC slab perturbed by the underlying InAs substrate—is numerically demonstrated using a combination of finite-element method (FEM) and finite-difference time-domain (FDTD) simulations. We design a structure (Fig. 2(a)) consisting of a 2.05-μm-thick dielectric PhC slab (dielectric constant $\varepsilon_r = 12.25$) on top of an InAs substrate. The PhC slab consists of a square lattice in the $xy$-plane (period 7 μm) of cylindrical through-holes (diameter 3.5 μm). Our design choice is partly motivated by the fact that the PhC slab is a canonical and well-understood structure in nanophotonics[28]. We used FEM simulations (COMSOL Multiphysics) to calculate the $s$-polarized spectral directional absorptance of the structure. The dielectric tensor of InAs is given by the gyrotropic Drude-Lorentz model:

$$\bar{\bar{\varepsilon}} = \begin{bmatrix} \varepsilon_\perp & 0 & +ig \\ 0 & \varepsilon_\parallel & 0 \\ -ig & 0 & \varepsilon_\perp \end{bmatrix}, \tag{2}$$

$$\varepsilon_\perp = \varepsilon_b - \frac{\omega_p^2(\omega + i\Gamma)}{\omega[(\omega + i\Gamma)^2 - \omega_c^2]}, \tag{3}$$

$$\varepsilon_\parallel = \varepsilon_b - \frac{\omega_p^2}{\omega(\omega + i\Gamma)}, \tag{4}$$

$$g = \frac{\omega_p^2\omega_c}{\omega[(\omega + i\Gamma)^2 - \omega_c^2]}, \tag{5}$$

where $\varepsilon_b$ is the background dielectric constant, $\omega_p = \sqrt{ne^2/(m^*\varepsilon_0)}$ is the plasma frequency, $\Gamma$ is the damping constant, $\omega_c = e|\boldsymbol{B}|/m^*$ is the cyclotron frequency, $n$ is the carrier concentration, $m^*$ is the electronic effective mass, $\varepsilon_0$ is the permittivity of free space, and $e$ is the elementary charge[29]. For the structure shown in Fig. 2(a), we assume $\varepsilon_b = 12.3$, $\omega_p = 3.80 \times 10^{14}$ rad/s, $\Gamma = 4.5 \times 10^{12}$ rad/s, and $m^* = 0.033m_e$ (borrowed from ref. 12), but the conclusions of our analysis are general. For simplicity, we also assume the PhC slab is lossless, but the InAs substrate is allowed to be lossy. In Fig. 2(b), the zero-field (i.e., $B = 0$) $s$-polarized spectral directional absorptance is plotted as a function of the longitudinal wavevector, $k_x = (\omega/c)\sin\theta$, normalized to the lattice vector $2\pi/a$, where $a$ is the period, and frequency, normalized to $c/a$. We observe a number of fairly dispersive absorption peaks, which suggests coupling to particular electromagnetic modes supported by the structure. We can, in fact, numerically solve for these electromagnetic modes and directly compare them to the spectral directional absorptance. The green points superimposed on the plot are the photonic band structure of the unperturbed PhC slab suspended in air, calculated using MIT Photonic Bands (MPB), a frequency-domain eigensolver[30]. The red points are the photonic band structure of the full, PhC-slab-on-InAs structure, calculated using harmonic inversion implemented in Meep[31–33]. These photonic band structures give us insight into the nature of the absorption peaks. We observe that (1) the red points agree with the absorption peaks, meaning $s$-polarized light couples to the TM eigenmodes, and (2) the green points follow the same trends as the red points, especially concavity and curvature. Near the ENZ wavelength, the effective refractive index of InAs starts to get relatively close to 1, the refractive index of air, suggesting the InAs substrate simply perturbs the dielectric environment of the PhC slab in air in this spectral range. This explains why the green and red points near the ENZ wavelength are so similar: they are the same set of TM eigenmodes, but one set is perturbed by the InAs substrate.

Our hypothesis that the InAs substrate perturbs the TM eigenmodes of the PhC slab near the ENZ wavelength is supported by Fig. 2(c), which shows mode profiles calculated using COMSOL Multiphysics. To do so, we launch an $s$-polarized plane wave toward the PhC slab, with and without the InAs substrate, with the aim of exciting the modes closest to the green and red arrows in Fig. 2(b). We observe similar, almost $C_4$-symmetric mode profiles in both structures, suggesting the absorption peak in the colormap corresponds to a perturbed TM eigenmode of the PhC slab indeed. The mode profiles are, in fact, $C_2$-symmetric because the nonzero value of $k_x$ breaks an additional symmetry in the system, but the eigenmode would be $C_4$-symmetric at the Γ-point, where $k_x = 0$. Thus, we argue that the absorption of $s$-polarized light in our design is mediated by these $xy$-polarized TM eigenmodes.

Although the above explanation focuses on $s$-polarized light, the nonreciprocity of $p$-polarized light is always present since it has components perpendicular to $\boldsymbol{B}$ in the Voigt configuration, regardless of the structure. In fact, the PhC slab can serve to control and enhance $p$-polarized nonreciprocity, e.g., via enhancement of the local density of optical states (LDOS), while enabling comparatively weaker $s$-polarized nonreciprocity. Furthermore, both $s$- and $p$-polarized light can excite $xy$-polarized TM eigenmodes, enhancing the polarization-averaged nonreciprocal response at the resonant frequencies of these modes. This is illustrated in Figs. 2(d–e). Figure 2(d) shows the spectral absorptance of $s$- (black) and $p$- (red) polarized light for forward- (solid) and backward- (dashed) propagating light at an angle of incidence of 50° and magnetic field of 1 T. Figure 2(e) shows the absorptance contrast, defined as the difference between forward absorptance $a_\ell(\omega, \theta, B)$ (solid curves) and the backward absorptance $a_\ell(\omega, -\theta, B)$ (dashed curves):

$$\Delta a_\ell(\omega, \theta, B) = a_\ell(\omega, -\theta, B) - a_\ell(\omega, \theta, B). \tag{6}$$

The blue curve is the polarization-averaged value, $\Delta a(\omega, \theta, B) = \left[\Delta a_s(\omega, \theta, B) + \Delta a_p(\omega, \theta, B)\right]/2$. At the peak and valley of $\Delta a(\omega, \theta, B)$, $\Delta a_s(\omega, \theta, B)$ and $\Delta a_p(\omega, \theta, B)$ have the same sign. As a result, the polarization-averaged peak and valley values are greater than half the corresponding value for either polarization alone. Therefore, the polarization-averaged nonreciprocity is stronger than if only one polarization were nonreciprocal. This points toward better utilization of nonreciprocal thermal emission and absorption compared to single-polarized designs. We remark that the ability to achieve nonreciprocity in both polarizations also points toward selectively tuning nonreciprocity in each polarization channel, designing a desired response for either polarization independently.

*Experimental demonstration of dual-polarized nonreciprocal absorption*

We fabricated a PhC-slab-on-InAs structure similar to the one shown in the preceding analysis but re-optimized for the InAs wafer used in our experiments. We measured the optical properties of the InAs wafer using ellipsometry and reflectance spectroscopy, finding that $\varepsilon_b = 12.243$, $\omega_p = 4.928 \times 10^{14}$ rad/s, $\Gamma = 4.556 \times 10^{12}$ rad/s, and $m^* = 0.0608 m_e$ (Eqs. (2–5)). The structure was fabricated by: (1) depositing a layer of a-Si via plasma-enhanced chemical vapor deposition (PECVD); (2) spinning, patterning, and developing a layer of electron beam resist; (3) depositing a layer of $Al_2O_3$ via electron-beam physical vapor deposition (PVD); (4) lift-off to create a hardmask; and (5) dry etching. A unit cell of structure is shown in Fig. 3(a). The period of the PhC slab is 6.105 μm, the hole diameter 4.053 μm, and the thickness 1.477 μm, as shown in Fig. 3(a). The dimensions were estimated using a combination of ellipsometry (to determine the thickness of the a-Si layer), scanning electron microscopy (SEM), and optical profilometry. SEM images are shown in Fig. 3(b). Overall, the holes appear to be of acceptable quality, though we observe some sidewall roughness that appears to be deeply subwavelength (tens to hundreds of nanometers, compared to mid-infrared wavelengths longer than 10 μm—at least two orders of magnitude difference). We confirm that the a-Si was etched all the way through using optical profilometry. Figure 3(c) shows an overlay of confocal and optical microscope images captured by the optical profilometer, with vertical and horizontal cuts indicated by the red and blue lines, respectively. These cuts are plotted in Figs. 3(c)(i–ii), showing that the hole depth is approximately 1.48 μm, consistent with post-PECVD ellipsometry.

Figure 3(d) shows good agreement between the experimental and simulated zero-field absorptance of the structure as a function of angle of incidence and wavelength, for propagation in the $xz$-plane. Each pair of heatmaps (one for each of the two polarizations) shows the experimentally measured and theoretically predicted values calculated using COMSOL Multiphysics. Figure 3(e) is the same as Fig. 3(d), but for propagation 45° from the $xz$-plane (i.e., along the diagonal of the unit cell), again showing good agreement between theory and experiments. This good agreement between theory and experiments along both azimuthal angles of incidence (0° and 45°, which map the high-symmetry lines of the square lattice) provides evidence of high fabrication fidelity to the designed structure. This is because the simulated zero-field absorptance takes in the independently measured dielectric functions of InAs and a-Si, as well as geometric parameters (period, hole radius, and a-Si thickness) extracted from optical profilometry and SEM images of a sample unit cell, without fitting to the experimental zero-field absorptance itself. In particular, good agreement between the apparent locations and bandwidths of resonant absorption peaks strongly suggests that the as-fabricated structure faithfully reproduces the desired optical response and that the sample unit cell is representative of the nearly 10 mm × 10 mm patterned area. Additionally, nonreciprocity is expected to be strongest in the vicinity of these absorption peaks, as they spectrally shift because of the magnetic field. For propagation in the $xz$-plane, which is the relevant configuration for our nonreciprocity experiments (to be discussed), we highlight resonant features at approximately 13 μm in the $p$-polarized absorptance, 13.5 μm in the $s$-polarized absorptance, and 11.5 μm for both polarizations. These features are indicated by the dotted lines on top of the experimental heatmaps in Fig. 3(d).

We then experimentally observed dual-polarized, nonreciprocal absorption at room temperature using a magnetic-field-integrated FTIR setup. The magnetic field was generated by a Halbach array. Its strength could be controlled by adjusting the gap width between the pole pieces, with narrower gaps having stronger magnetic fields. To align the sample *in situ* with the heavy Halbach array, we built a custom, 5-axis stage to control the $z$-position, height, and $x$- and $y$-tilt of the sample. A photograph of the stage is shown in Fig. 4(a). We also mapped the magnetic field as a function of position in the $xy$-plane (setup shown in Supplementary Fig. S5). An example magnetic field map corresponding to the conditions of our nonreciprocal absorption experiments is shown in Fig. 4(a)(i), along with a simulated map for comparison. For the gap width of interest in this work, the spatially-averaged applied magnetic field was approximately 1 T. Additional magnetic field maps and the spatially-averaged magnetic field as a function of gap width are shown in Supplementary Figs. S6 and S7, respectively.

A schematic of the sample between the pole pieces and the plane of incidence is shown in Fig. 4(b). Our goal is to measure the absorptance contrast, Eq. (6), and experimentally demonstrate that it is nonzero. In our experiments, we used a commercially available infrared ellipsometer (see Methods), so it was impossible to reverse the light propagation direction, which would have provided the direct positive to negative angle of incidence delta signifying nonreciprocity. Instead, we invoked the Onsager-Casimir reciprocal relations, which establish the equivalence between a reversal of the magnetic field and reversal of light propagation[34,35]. In other words,

$$a_\ell(\omega,-\theta,B) - a_\ell(\omega,\theta,B) = a_\ell(\omega,\theta,-B) - a_\ell(\omega,\theta,B). \quad (7)$$

This is relatively straightforward to prove via symmetry arguments and has been experimentally demonstrated for thermal emission and absorption[8].

Figure 4(c) shows that we have successfully demonstrated dual-polarized nonreciprocal absorptance. Experimental and simulated heatmaps of the (i) $p$- and (ii) $s$-polarized absorptance contrasts, which are simultaneously nonzero in the 11–12 μm and 13.5–14 μm spectral ranges within the ENZ region, where the resonances in Fig. 3(d) are located. Again, good agreement between theory and experiments affirms the fidelity of our sample fabrication, magnetic field measurements, and experimental setup and methodology. One notable difference is the slightly nonzero absorptance contrast in the 14–18 μm spectral range observed in experiments, which in theory should be closer to 0. To understand where this "background absorptance contrast" might come from, recall that asymmetry in absorption is only a signature of nonreciprocity in planar, subwavelength systems, as opposed to diffractive systems. In the 14–18 μm spectral range, however, it is possible that unintended scattering and diffraction give rise to apparently broadband, nonresonant, nonzero absorptance contrast (discussed in detail in Sec. IV of the Supporting Information). On the other hand, we see a comparatively stronger nonzero absorptance contrast at wavelengths that line up with the resonant absorption peaks in Fig. 3(d) (indicated by the dotted lines). These resonances are subwavelength in origin: they are a manifestation of the Bloch eigenmodes in the PhC slab. This means that the corresponding nonzero values of $\Delta a_{\ell}(\omega, \theta, B)$ can be ascribed to nonreciprocity rather than asymmetric diffraction.

Figure 4(d) provides further confirmation that the peaks and valleys in absorptance contrast originate from the spectral shifting of resonant absorption peaks due to the applied magnetic field. It shows the magnetic field-dependent absorptance of (i) $p$- and (ii) $s$-polarized light and vertical cuts of the absorptance contrast from 11–14 μm (the approximate ENZ region) at $\theta = 35°$. For the absorptance, theory (lines) accurately captures the general trends observed in experiments (markers), with the zero-field spectrum (blue) lying between the $-B$ (yellow) and $+B$ (orange) spectra. This plot clearly shows that nonreciprocity arises from resonant absorption peaks spectrally shifting from the central zero-field spectrum in response to the applied magnetic field, once again highlighting peaks at approximately 11.5 and 13 μm for $p$-polarization and 11.5 and 13.5 μm for $s$-polarization. We note, however, that these peaks are slightly blueshifted and differ in magnitude in experiments compared to theory. We attribute these minor discrepancies to imperfections in both fabrication and the experimental setup, discussed in detail in Sec. IV of the Supporting Information, with additional evidence from FEM simulations.

**Discussion**

In this work, we observed a dual-polarized, nonreciprocal optical response in the 11–14 μm spectral range, close to the peak wavelength of room-temperature thermal radiation (approximately 10 μm). Our design was based on a PhC slab supporting Bloch eigenmodes with components perpendicular to the magnetic field that were accessible via *s*-polarized light, enabling magneto-optical near-field interactions via mode overlap. The outcome is that both *s*- and *p*-polarized light showed nonreciprocal absorptance, in reasonably good agreement with theoretical predictions. Our design is relatively simple, with few design parameters, and based on a canonical, well-studied photonic structure whose resonances can be tuned by varying those few design parameters (e.g., period, hole diameter, thickness).

For the family of photonic crystal slabs we studied, one might ask how the *s*-polarized nonreciprocal optical response can be strengthened, as it is relatively weak compared to *p*-polarized. One approach is to operate off-Voigt configuration, meaning the applied magnetic field possesses components parallel and perpendicular to the plane of incidence. In this situation, *s*-polarized light would have a fundamentally nonreciprocal optical response since $\boldsymbol{E}_s$ would always have components perpendicular to $\boldsymbol{B}$, but there are two drawbacks of this approach. Firstly, light propagating inside InAs would no longer be *s*- and *p*-polarized. Instead, it would take on a more general, elliptical polarization. As a result, linearly polarized light incident from air would experience polarization conversion upon reflection due to the boundary conditions of Maxwell's equations coupling electric and magnetic field components across the interface between air and InAs. This polarization conversion manifests in the reflection matrix having $s$ to $p$ and $p$ to $s$ components, $R_{sp}$ and $R_{ps}$. This would complicate the analysis and design of a structure to enhance $s$-polarized nonreciprocal absorption. More importantly, non-Voigt configurations have an overall weaker nonreciprocal optical response due to having higher spatial symmetries compared to Voigt. In the limit that the magnetic field is parallel to the plane of incidence, known as the Faraday configuration, reflectance and absorptance become fully reciprocal due to inversion symmetry[36]. Therefore, there is no obvious solution that makes use of some sort of non-Voigt configuration, as the relative strengths of these trade-offs are not immediately clear, in part due to the complexity of the problem involving elliptical polarization. For future designs, this does not rule out the utility of non-Voigt configurations, only that designing structures that exploit the benefits of these configurations may not be so straightforward.

Within the scope of the more design-friendly Voigt configuration, loss is arguably the most important factor affecting the performance of nonreciprocal absorbers, dual-polarized or not, creating a need for better mid-infrared gyrotropic materials. Similar to existing designs based on GMRs, nonreciprocal absorption in our design relies on shifting the ENZ frequency in response to $\boldsymbol{B}$ and the resulting shifts of resonant absorption peaks. For this reason, the quality factors (Q-factors) of these resonances play an important role: the greater the Q-factor, the narrower the peak and, therefore, the stronger the absorption contrast when the peak spectrally shifts because of the applied magnetic field. High-Q-factor GMRs in nonreciprocal systems have been postulated to enable a near-total breakdown of Kirchhoff's law of thermal radiation—meaning almost 100% absorption and 0% emission for a given frequency, direction, and polarization, and vice versa for the opposite direction[11,37]. In dielectric structures, GMRs, bound states in the continuum (BICs), Bloch eigenmodes, and other nanophotonic resonances can have high Q-factors, but when InAs is introduced into these structures, losses can significantly broaden otherwise sharp resonant absorption peaks. Here, "losses" include scattering due to fabrication defects as well as intrinsic, material loss, which can be relatively large in doped semiconductors such as InAs and only worsens as the carrier concentration increases. This implies a fundamental trade-off between the spectral range of the ENZ region, where the nonreciprocal optical response is the strongest, and loss—in other words, a trade-off between the desired frequency and imposed bandwidth of the absorption peak. Longer ENZ wavelengths that go into the far-IR benefit from higher-Q resonances, but the power carried by thermal emission at temperatures relevant to practical applications (at least room temperature) is relatively low in this band, limiting the usefulness of the stronger emission-absorption contrast afforded by lower loss. Ultimately, due to these material-level trade-offs and the limiting effect of loss on absorptance contrast, there is a need to seek out alternative mid-infrared, gyrotropic materials that may circumvent the challenges associated with InAs and related semiconductors.

Moreover, the dielectric resonator material (in our case a-Si) also exhibits loss at mid-infrared wavelengths, which also decreases the Q-factor but can be addressed through more careful fabrication or different design choices. PECVD-deposited a-Si has an absorption peak at 15.5 μm, which is unfavorable in the mid-infrared spectral range regardless of the doping level of InAs[38]. This absorption peak is associated with Si-H bonding; hence, it could be mitigated by more carefully controlling the deposition parameters, using a different deposition method free of hydrogen or other impurities such as PVD, or using an entirely different dielectric material depending on the targeted spectral range. The challenge of loss in the PhC slab itself (rather than the substrate it lies on top of) can also be addressed designing the structure to operate farther away from a-Si's absorption peak or using an entirely different dielectric material such as Ge.

Finally, we point toward strong electric field enhancement as an established approach to increasing Q-factor in low-loss resonators such as dielectric PhC slabs[39]. The mode profiles shown in Fig. 2(c) are normalized to the electric field magnitude of the incident wave in our FEM simulations, meaning the color of the heatmaps indicates the electric field enhancement provided by the structures. The PhC slab in air (Fig. 2(c)(i)) can achieve a nearly 10-fold enhancement, while the PhC slab on InAs (Fig. 2(c)(ii)) is limited to an approximately 2-fold enhancement. This suggests there is significant room for improvement through designing for electric field enhancement, for example by using a back reflector to confine the electric field as many theoretical proposals for strong nonreciprocity do[11,23–26,37].

In addition to Q-factor engineering, nonreciprocal absorption can be strengthened by decreasing $m^*$, the electronic effective mass in Eq. (5). For a given $B$-value, the magnitude of the off-diagonal component $g$ of the dielectric tensor can be increased by decreasing $m^*$, via the fact that $\omega_c = eB/m^*$. For this reason, it has been suggested that InSb would be superior to InAs for nonreciprocal absorption because of its comparable mobility and lower effective mass[40], but it is more difficult to source. Nevertheless, we believe InSb is a promising alternative to InAs for nonreciprocal thermal emitters and absorbers.

Finally, in this work, we report an experimental demonstration of a 2D photonic crystal applied to mid-infrared absorption at thermal wavelengths[27]. Photonic crystals of different dimensionalities (1D, 2D, and/or 3D) and symmetries can be used to tailor thermal emission and absorption, enabling spectral selectivity[41,42], photonic band gaps[43–47], coherence[48], and nontrivial photonic band topology[49–51] in the mid-infrared spectral range. In such systems, an applied magnetic field could act as a time reversal symmetry-breaking perturbation, effectively making the photonic band structure asymmetric with respect to the wavevector $\boldsymbol{k}$ (i.e., nonreciprocal), even making certain eigenmodes $(\omega, \boldsymbol{k})$ accessible in one propagation direction inaccessible in the opposite one, $(\omega, -\boldsymbol{k})$. At the same time, nonreciprocity as an optical phenomenon is inherently weak, constrained by material selection and strongly dependent on wavelength, direction, and polarization. By contrast, thermal emission is broadband, broad angle, and unpolarized, leading to a fundamental gap in the field of nonreciprocal thermal photonics between the thermal emission channels available for nonreciprocal modulation and what existing nonreciprocal physical mechanisms, materials, and designs can achieve. However, since photonic crystals offer selectivity over spectrum, direction, polarization, and LDOS, they could help bridge the aforementioned gap between the fundamental limitations of nonreciprocity and opportunities afforded by the availability of many spectral and directional channels in thermal emission. Provided that the longstanding challenge of loss can be overcome, simple photonic crystal-based design motifs such as the one presented here can be further adapted to nonreciprocal systems, with the additional degree of freedom of decoupled thermal emission and absorption these systems offer. In turn, more advanced control enabled by nonreciprocal photonic crystals can impact applications such as energy conversion, optical isolation and circulation in the mid-infrared, thermal camouflage, and thermal management.

## Methods

*Fabrication*

The a-Si PhC slab was fabricated on top of an n-type, Sn-doped, (100)-oriented InAs wafer sourced from MTI Corporation. The dimensions of the wafer were 10 mm × 10 mm × 0.5 mm. a-Si was deposited using PECVD (Oxford Instruments Plasmalab System 100) with 5% $SiH_4$ in $N_2$ at 200 °C. The deposition rate was approximately 24 nm/min. Prior to depositing the a-Si film on top of the InAs wafer, we conditioned the chamber by running the process for 10 minutes on the carrier wafer (without the sample). After depositing a-Si, we spin-coated negative electron beam resist (ma-N 2403, micro resist technologies) on top of the sample using a Laurell WS-650. The sample was spun at 3000 rpm for 30 s, followed by baking on a 90 °C hotplate for 60 s. The pattern was prepared using the KLayout and BEAMER (GenISys GmbH) software, then the patterns were written using a Raith EBPG 5000+ electron beam pattern generator. The current was 100 nA, voltage 100 kV, dose 400 $\mu C/cm^2$, and aperture size 300 μm. After writing, the resist was developed by submerging in MF-319 developer (DuPont) for 70 s, then rinsed by gently flowing DI water over it for 1 min, leaving behind pillars of developed resist. Then, an electron-beam evaporator (Angstrom Engineering Amod) was used to deposit approximately 50 nm of $Al_2O_3$ on top of the a-Si and developed resist. This thickness of $Al_2O_3$ was chosen because it offers good etch selectivity while being optically thin in the mid-infrared (despite its strong absorption peak around 14 μm[52]; the negligible effect of $Al_2O_3$ on the absorptance was validated through numerical simulations, see Supplementary Fig. S8). To create the holes in the hardmask to be etched, we constructed a water bath to heat Remover PG (Kayaku Advanced Materials) to 60 °C, removing it from the heat, then leaving the sample submerged in it overnight to lift off the developed resist. Then, the Remover PG with the sample in it was placed in an ultrasonic bath and sonicated at room temperature for 5 min, followed by additional sonication in acetone and isopropyl alcohol for 5 min each. This fully removed the swollen resist, leaving behind a holey $Al_2O_3$ hardmask. Finally, the pattern was etched through the a-Si using an inductively coupled plasma reactive-ion etcher (Oxford Instruments Plasmalab System 100 ICP 380). We used the Bosch process to achieve anisotropic dry etching, with an etch rate of approximately 482.8 nm/min. Before etching the pattern, we conditioned the chamber by running the process for 10-minutes on the carrier wafer. To minimize overetching, we set the etch time to the a-Si thickness (measured using ellipsometry) divided by the etch rate, plus a few seconds to make sure the a-Si was etched all the way through. A 20-minute $SF_6/O_2$ cleaning process was run before and after etching. The etch depth, intended to be equal to the a-Si film thickness, was confirmed using optical profilometry (Keyence VK-X3000, see Fig. 3(c)). Additional imaging and characterization were performed using optical microscopy (Keyence VHX-7000) and SEM (FEI Sirion, see Fig. 3(b)). The fabrication process is illustrated in Supplementary Fig. S1, with optical microscope images after different steps shown in Supplementary Fig. S2.

*Ellipsometry*

The optical properties of the InAs wafer were measured at zero applied magnetic field using an infrared ellipsometer (IR-VASE Mark II, J.A. Woollam). The dielectric function was fitted to the Drude model using WVASE, the software included with the ellipsometer:

$$\varepsilon_{\mathrm{InAs}}(\omega) = \varepsilon_{b,\mathrm{InAs}} - \frac{\omega_p^2}{\omega(\omega + i\Gamma)}. \tag{8}$$

After depositing a-Si, the thickness and optical properties of the layer were fitted assuming we knew the dielectric function of InAs, from fitting Eq. (8). The dielectric function of a-Si was assumed to be of the form:

$$\varepsilon_{\text{a-Si}}(E) = \varepsilon_{b,\text{a-Si}} + \frac{A\gamma\omega_0}{\omega_0^2 - \omega^2 - i\gamma\omega}, \quad (9)$$

where $\varepsilon_{b,\text{a-Si}}$ is the background dielectric constant of a-Si, $A$ is the oscillator strength, $\gamma$ is the damping constant, and $\omega_0$ is the center frequency. This models the absorption peak around 15.5 μm as a Lorentz oscillator. For the sample shown in this work, we found that $\varepsilon_{b,\text{a-Si}} = 11.115$, $A = 2.777$, $\omega_0 = 1.203 \times 10^{14}$ rad/s, and $\gamma = 1.570 \times 10^{13}$. After depositing the hardmask, additional measurements were carried out on an unreported sample with an unpatterned, multilayer region (InAs/a-Si/$Al_2O_3$) to confirm that the hardmask had a negligible effect on the optical response of the structure. The measured and fitted ellipsometry data are shown in Supplementary Figs. S3–S4, along with the dielectric functions following Eqs. (8–9).

*Magnetic-field-integrated absorptance spectroscopy*

The applied magnetic field was generated by a Halbach array, purchased from the research group of Dr. Arne Laucht at the University of New South Wales[53]. This exact Halbach array was used in prior work[8,12,14]. It consists of neodymium magnets with a pair of Supermendur pole pieces, held together by copper housing. This arrangement of permanent magnets generates a fairly uniform magnetic field distribution at its center, and the magnetic field strength can be tuned by moving the pole pieces using built-in screw threads. We 3D printed (Bambu Lab X1 Carbon) a set of shims to precisely and repeatably control the gap width between the pole pieces. The magnetic field distribution was measured as a function of said gap width using a Hall probe connected to a handheld teslameter (AMTAST TES11A). The Hall probe was held in place by a custom, 3D-printed holder and a set of optical posts, which were mounted on a translation stage and a lab jack, providing full range of translation in the $xy$-plane (Supplementary Fig. S5). These measurements allowed us to estimate the magnetic field acting on the sample during absorption spectroscopy. We confirmed the accuracy of the measured magnetic field distributions by simulating the Halbach array using Radia, a Python implementation of the boundary element method[54,55]. A comparison of a measured magnetic field distribution and a simulated one is shown in Fig. 4(a)(i), with additional results shown in Supplementary Fig. S6. Spatially-averaged values as a function of gap width are shown in Supplementary Fig. S7; we set the gap width to 11 mm, resulting in an approximately 1 T applied magnetic field. We observe a slight difference between the magnetic field magnitudes in experiments and simulations, with the Radia-simulated magnetic field being slightly stronger. This difference could be due to the gradual demagnetization of the permanent magnets in the Halbach array over time, as they are several years old. The simulations assumed the permanent magnets were N52 neodymium magnets of remanent magnetization 1.45 T, but in the experiments, the permanent magnets could have been (and likely were) weaker[56].

The same ellipsometer was used in its reflectance mode to measure the frequency, angle, and polarization-dependent reflectance of the sample. Since the Halbach array was heavy (over 3.5 kg) and did not fit on the sample stage included with the ellipsometer, we built a custom, 5-axis stage to enable translational and rotational alignment of both the sample and the Halbach array. The 5-axis stage consisted of a linear $z$-axis translation stage (Thorlabs PT1B), a linear $y$-axis translation stage (Newport DS65-Z), a goniometer for tilt about the $x$-axis (Melles Griot), a rotation stage for tilt about the $y$-axis (Thorlabs XRNR1), and a 3D printed pin-and-slot mechanism for linear $x$-axis translation. We also 3D printed adapters to attach each stage to the one below it, as not all of them were compatible (e.g., the goniometer had metric threads and the other stages imperial). Two different adapters were 3D printed to mount the Halbach array right side up and upside down, enabling measurements as a function of magnetic field direction or, equivalently, light propagation direction via the Onsager-Casimir reciprocal relations[34,35].

Since the sample was optically thick and opaque, absorptance was related to reflectance by $a_\ell(\omega,\theta,B) = 1 - R_\ell(\omega,\theta,B)$. To obtain a reference, near-100% reflectance signal that accounts for the shape and divergence of the beam and ambient environment, we measured the intensity of light reflected off an Au mirror. Then, the reflectance of the sample was given by:

$$R_\ell(\omega,\theta,B) = \frac{I_{\text{sample}}(\omega,\theta,B)}{I_{\text{mirror}}(\omega,\theta,B)}. \tag{10}$$

We assumed that the reflectance of the mirror is unaffected by the magnetic field, as Au is not known to support any magneto-optical effects in the mid-infrared spectral range, nor did we experimentally observe any evidence of this. However, to most accurately calculate the reflectance, the ambient environment surrounding the sample and the mirror (which can affect the background signal or cause scattering) needed to be as close to identical as possible, even when changing the magnitude and direction of $B$. To achieve this similitude across experimental conditions, we fabricated the mirror on top of a glass substrate (VWR 16004-368) the same size as the sample (10 mm × 10 mm lateral area). It consisted of 5 nm of Ti for adhesion and 100 nm of Au, deposited using the same electron-beam evaporator as before. The maximum beam spot size was 10 mm × 5 mm at $\theta = 60°$ (where it varies as $5/\cos\theta$ in units of mm), so at worst, slight apodization by the sample was expected. Nevertheless, a mirror of the same size as the sample, in addition to masking the sample and the mirror with carbon tape (Ted Pella 16073-4), allowed for this apodization to be accounted for in principle. We measured the mirror directly after measuring the sample in each experimental condition so that any apodization, scattering, or background signal due to the sample stage itself is also accounted for. Finally, to ensure that even for the zero-field absorptance shown in Fig. 4(d) the ambient environment was similar, we 3D printed a "dummy" Halbach array and used aluminum foil and copper tape to create an optical facsimile of the real Halbach array. Photographs of both Halbach arrays are shown in Supplementary Fig. S12. We note, however, that the zero-field absorptance shown in Fig. 3(d–e) was measured using the sample stage included with the ellipsometer. Due to the dimensions of the sample, it would not have been possible to measure absorptance along the diagonal of the sample (i.e., azimuthal angle of incidence 45°) between the pole pieces, whereas the included sample stage is more open. As a result, the magnitude of the zero-field absorptance in Figs. 3(d–e) is slightly different from that in Fig. 4(d).

The ellipsometer facilitated the measurement of absorptance automatically as a function of frequency $\omega$ (or wavelength $\lambda$), polar angle of incidence $\theta$ (via a motorized 2 theta stage), and both input and output polarization. We measured the polarization-dependent, spectral directional absorptance from 35° to 60° in 1° increments, with 40 scans per spectrum (i.e., each spectrum was the average of 40 repeated measurements). This meant measurements were slow (taking up to three hours per $B$-value) but had relatively high resolution in $\omega$-$\theta$ space for comparison with theory.

Finally, from the ellipsometry described in the previous section, it was not possible to fit the electronic effective mass $m^*$ of InAs. This quantity can only be estimated by measuring the optical response (ellipsometric $\Psi$ and $\Delta$ or absorptance) as a function of magnetic field[15]. We measured the magnetic field-dependent absorptance of a similar, pristine InAs wafer and estimated $m^*$ to be approximately $0.0608m_e$, where $m_e$ is the electron mass. This value produces good agreement between theory and experiments.

## Frequency-domain modeling

### *COMSOL Multiphysics*

Polarization-dependent, spectral directional absorptance was calculated via FEM simulations in COMSOL Multiphysics with the Wave Optics Module. The optical properties of InAs and a-Si, measured via ellipsometry, were imported as analytical dielectric functions, following Eqs. (2–5), (8), and (9). The dimensions of the sample measured using a combination of ellipsometry, optical microscopy, and SEM were used to construct the geometry. The simulation domain consisted of periodic boundary conditions in $x$ and $y$ (consistent with the coordinate systems shown in all the figures), superimposed periodic ports for *s*- and *p*-polarized light, and a perfectly matched layer (PML) below the InAs to treat the wafer as semi-infinite (especially important for wavelengths shorter than the ENZ wavelength, where InAs is more dielectric than metallic). Automatic meshing was performed by COMSOL Multiphysics and found to be sufficient. The mode profiles shown in Fig. 2(c) were normalized to the electric field magnitude of the incident wave $|\boldsymbol{E}_{inc}| = \sqrt{2Z_0 P/(A\cos\theta)}$, where $Z_0$ is the impedance of free space, $P = 1$ W is the power input, $A$ is the area of the periodic port (i.e., the period of the structure squared), and $\theta$ is the polar angle of incidence.

*MPB*

To calculate the photonic band structure of the isolated, PhC slab in air (green points in Fig. 2(b)), we used MPB, a direct, frequency-domain eigenmode solver[30]. MPB assumes periodic boundary conditions in all directions, so to isolate the PhC slab in the *z*-direction, we defined a supercell (approximately 4 periods tall; we found this to be sufficient without introducing additional Fabry-Pérot oscillations that would be valid solutions for thicker supercells). We solved for 16 TE and TM bands, with a resolution of 32 pixels per unit cell and 49 k-points.

**Time-domain modeling**

We used harmonic inversion to calculate the photonic band structure of the full, PhC-slab-on-InAs structure. This method is necessary because frequency-domain methods, such as those implemented in COMSOL Multiphysics and MPB, fail for dispersive media. In Meep, we excite the system in the time domain using broadband, randomly-placed sources, then use harmonic inversion, built into Meep, to extract the spatial and frequency eigenmodes of the system[31–33]. The simulation domain was similar to the previously mentioned models, but we added PMLs to the top and bottom. These PMLs were quite thick (8 unit cells) to kill the periodic boundary condition in the $z$-direction as much as possible. We set the resolution to 32 pixels per unit cell, resolving the minimum skin depth of InAs with 5 pixels, and set the Courant parameter to 0.2. To excite the system, we randomly placed 10 wide-bandwidth ($\Delta\omega = 1.0(c/a)$) Gaussian pulse sources in the simulation domain. This helped ensure the eigenmodes of the system were consistently excited, and the electric field of the source determined the polarization of the eigenmodes we excited. To excite a particular eigenmode, e.g., to plot the mode profile, we drastically narrowed the bandwidth ($\Delta\omega < 0.1$) but kept everything else the same.

**Acknowledgements:** This work was supported by the Space Solar Power Project. The authors thank Andrew Nyholm for helping fabricate the gold mirrors and supervising Y.Z. during parts of his fellowship. The authors thank Jared Sisler and Martin Thomaschewski for helpful discussions on fabrication, as well as Sachin Vaidya and Jinseok Kong for advice on simulations. The authors also thank Guy DeRose, Nathan Lee, Kelly McKenzie, Yonghwi Kim, Ivy Chen, and Fabian Williams for training on the deposition, lithography, and characterization tools critical to this work. S.P. gratefully acknowledges the support of the Resnick Sustainability Institute at Caltech via the Pioneer Postdoctoral Fellowship. Y.Z. was supported by a Summer Undergraduate Research Fellowship (SURF) via Caltech. We gratefully acknowledge the technical support and infrastructure provided by the Kavli Nanoscience Institute at Caltech.

**Author contributions:** S.P. and H.A.A. originally conceived the project. S.P. carried out FDTD and FEM simulations with input from R.S. and Y.-C.L. S.P. and Y.Z. built the magnetic-field-integrated absorptance spectroscopy setup and performed experiments. Y.Z. measured and simulated the magnetic field distribution generated by the Halbach array. S.P. and Y.-C.L. developed the fabrication process for the photonic crystal slab. S.P. fabricated the gold mirrors. S.P. performed imaging and characterization of the sample. All authors contributed to the writing and preparation of this manuscript.

**Data availability:** The data supporting the findings of this study can be found in the main text and Supplementary Information. Additional data, code, and CAD models for 3D printed parts may be available from the corresponding authors upon reasonable request.

**Conflict of interest:** The authors declare no competing interests.

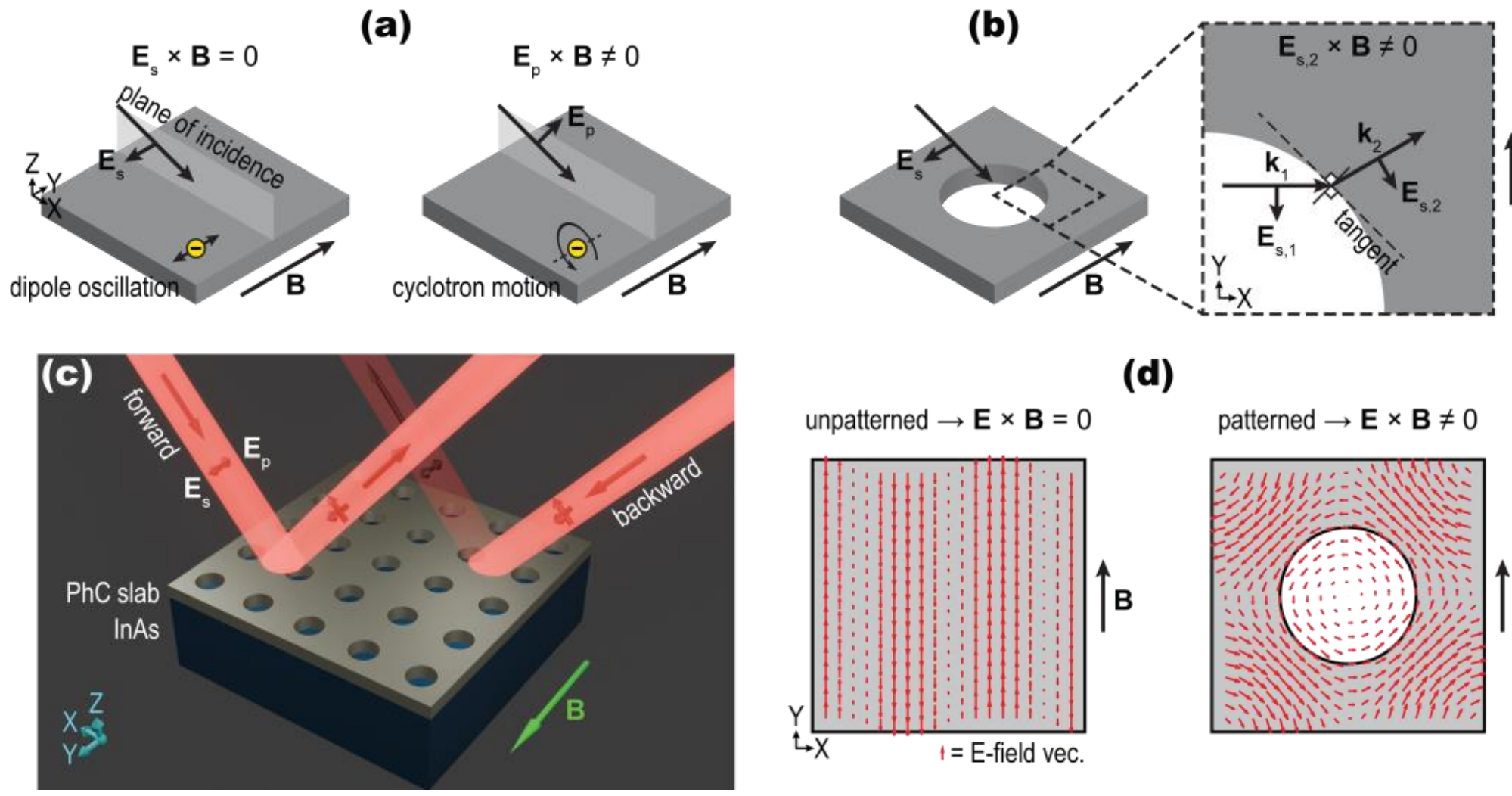

**Figure 1: Physical intuition behind $s$-polarized reciprocity and nonreciprocity.** (a) $s$-polarized light is fundamentally reciprocal because its electric field, $\boldsymbol{E_s}$, is parallel to the applied magnetic field $\boldsymbol{B} = B\hat{\boldsymbol{y}}$, which means that the induced dipole oscillation does not experience the Lorentz force. $\boldsymbol{E_p}$ is perpendicular to $\boldsymbol{B}$, resulting in a nonreciprocal, magneto-optical light-matter interaction via time reversal symmetry-breaking cyclotron motion. (b) Introducing holes forces $s$-polarized light entering the PhC slab, $\boldsymbol{E_{s1}}$ (wavevector $\boldsymbol{k_1}$) to acquire components perpendicular to $\boldsymbol{B}$ when transmitted, $\boldsymbol{E_{s2}}$ (wavevector $\boldsymbol{k_2}$), enabling nonreciprocity. (c) Dual-polarized nonreciprocal absorber concept, consisting of a subwavelength photonic crystal (PhC) slab on top of a gyrotropic material such as doped InAs. In the presence of $\boldsymbol{B}$ (green arrow), both $p$- and $s$-polarized light (indicated by the orthogonal vectors $\boldsymbol{E_s}$ and $\boldsymbol{E_p}$) may be efficiently reflected when propagating in the forward direction (left-to-right), but not in the backward direction (right-to-left), illustrated by the dimmer beam. (d) Patterning breaks continuous translational symmetry and introduces, e.g., $C_4$ symmetry. Instead of plane waves (left), the solutions of Maxwell's equations are $C_4$-symmetric Bloch eigenmodes, which can be $xy$-polarized (right, calculated using COMSOL Multiphysics). The coordinate system is shown in the bottom left of (a) and is consistent across the subfigures (a,b,d).

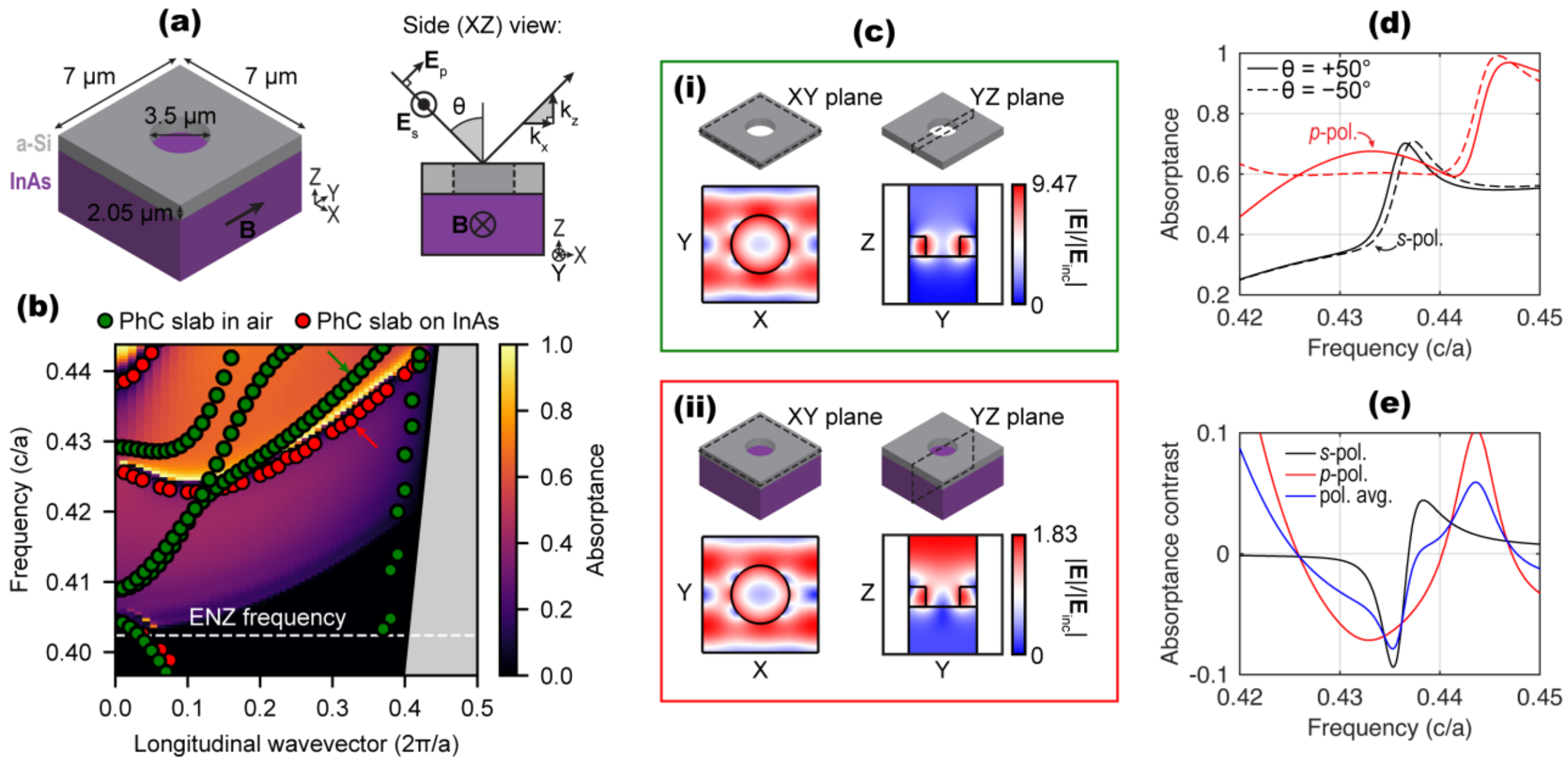

**Figure 2: Design of a dual-polarized nonreciprocal thermal absorber.** (a) Example of a dual-polarized nonreciprocal absorber with an InAs substrate of ENZ wavelength 17.4 μm. The period of the PhC slab is 7 μm, the hole diameter 3.5 μm, and the thickness 2.05 μm (the substrate is semi-infinite). (b) (Heatmap) Zero-field *s*-polarized absorptance. (Points) Photonic band diagram of the in-plane (TM) modes of the PhC slab in air (green) and on an InAs substrate (red). The dashed white line indicates the ENZ frequency, $\omega_p/\sqrt{\varepsilon_b}$. (c)(i) Mode profiles of the PhC slab in air, for the mode indicated by the green arrow in (b). These plots show the magnitude of the electric field $|\boldsymbol{E}|$ normalized to that of the incident wave, $|\boldsymbol{E}_{inc}|$. (ii) Same as (i), but for the mode indicated by the red arrow in (b). (d) *s*- (black) and *p*- (red) polarized absorptance for forward and backward angles of incidence, $\theta = \pm 50°$ and $|\boldsymbol{B}| = 1$ T. Both polarizations are nonreciprocal. (e) The difference between the solid and dashed lines in (d), or absorptance contrast. A nonzero value indicates nonreciprocity. The polarization-averaged value is shown in blue.

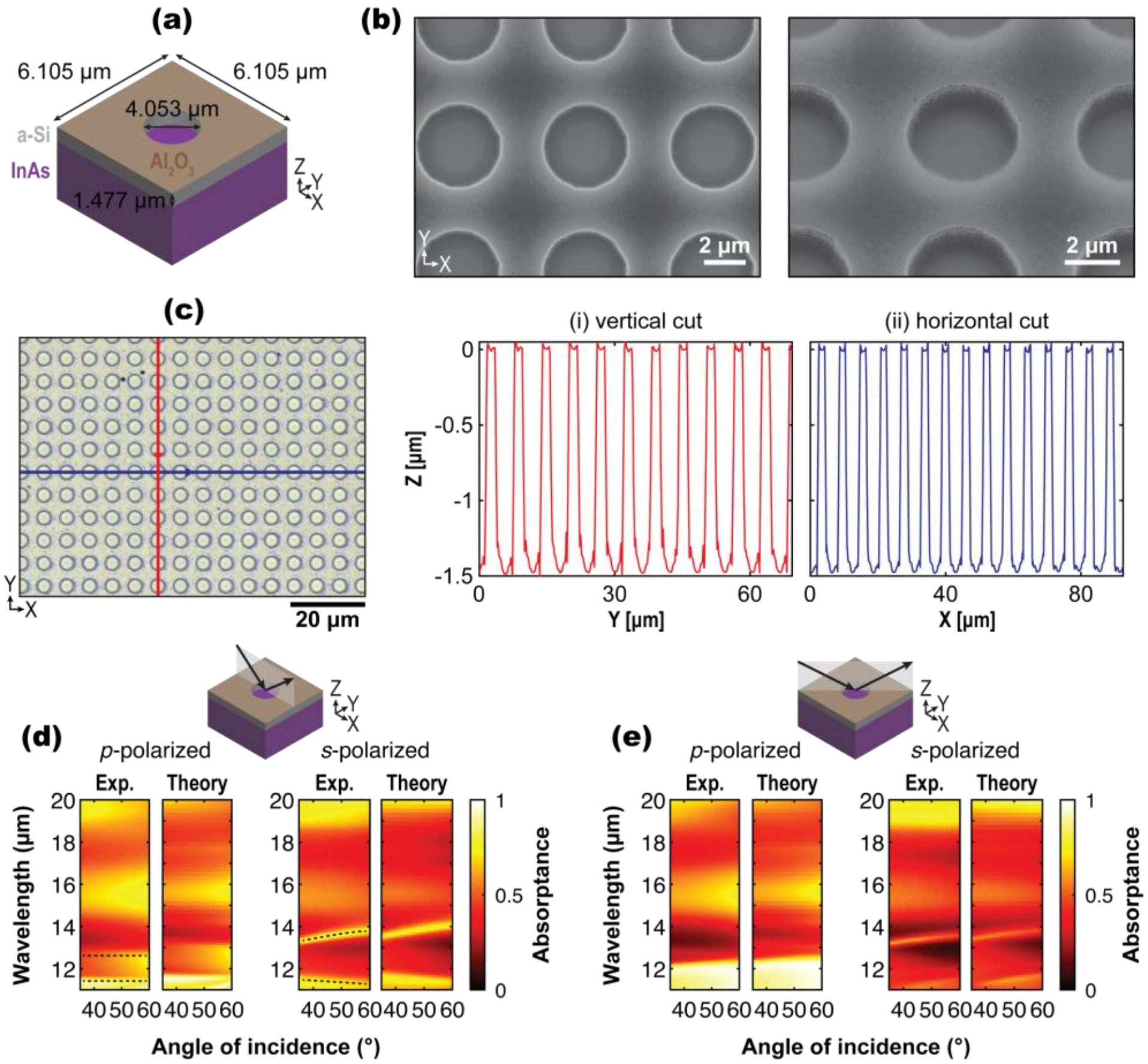

**Figure 3: As-fabricated dual-polarized nonreciprocal absorber.** (a) Dimensions of the structure, consisting of an opaque, 500-μm-thick InAs wafer (ENZ wavelength 13.2 μm), a-Si PhC slab, and $Al_2O_3$ hardmask, which is optically thin. The period of the PhC slab is 6.105 μm, the hole diameter 4.053 μm, and the thickness 1.477 μm. (b) SEM images of the structure, viewed from the top ($xy$-plane) and at an angle. (c) Optical profilometry, showing that the holes are etched all the way through the a-Si. (i) and (ii) are horizontal and vertical cuts, indicated by the red and blue lines in the overlaid confocal and optical microscope images. (d) Zero-field absorptance of the structure for light propagating in the $xz$-plane, showing good agreement between experiments and theory (calculated using COMSOL Multiphysics). Dotted lines on top of the experimental heatmaps are hand-drawn guides to the eye to highlight resonant features. (e) Zero-field absorptance for light propagating 45° from the $xz$-plane. These absorptances were measured using the sample stage included with the ellipsometer.

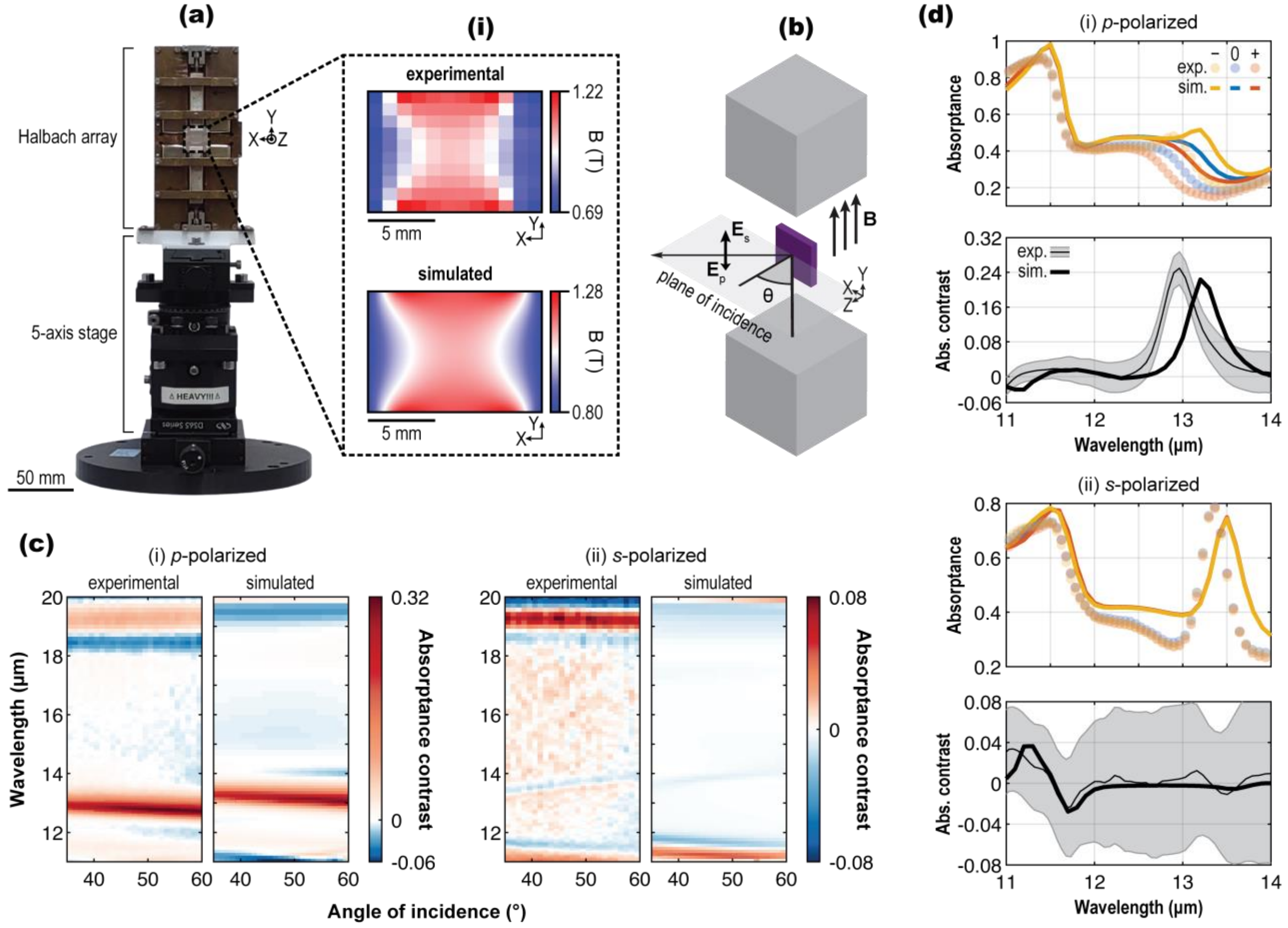

**Figure 4: Experimental observation of dual-polarized nonreciprocal absorptance.** (a) 5-axis, magnetic-field-integrated sample stage. (i) shows the magnetic field distribution generated by the Halbach array, comparing experimental and simulated magnetic field maps. (b) Schematic of the sample between the pole pieces (grey cubes) with the plane of incidence and polarizations, $s$ and $p$ ($s$ lies in the $xy$-plane, while $p$ lies in the $xz$-plane). (c) Experimental and simulated spectral directional absorptance contrast for (i) $p$- and (ii) $s$-polarized light and $B = 1$ T. (d) Magnetic field-dependent absorptance and partial line cut of absorptance contrast at $\theta = 35°$. In the absorptance plots, experimental data are the markers, while simulated data are the lines; different colors indicate negative (yellow), positive (orange), and zero (blue) magnetic field. Here, the zero-field absorptance was measured using a "dummy" Halbach array (Supplementary Fig. S12). In the absorptance contrast plots, experimental data are the thinner lines with shaded areas to represent error bars (see the Supplementary Information for uncertainty propagation), while simulated data are the thicker lines.

## References

1 Fan S. Thermal Photonics and Energy Applications. *Joule* 2017; **1**: 264–273.

2 Kirchhoff G. Ueber das Verhältniss zwischen dem Emissionsvermögen und dem Absorptionsvermögen der Körper für Wärme und Licht. *Annalen der Physik und Chemie* 1860; **185**: 275–301.

3 Luo C, Narayanaswamy A, Chen G, Joannopoulos JD. Thermal Radiation from Photonic Crystals: A Direct Calculation. *Physical Review Letters* 2004; **93**: 213905.

4 Chan DLC, Soljačić M, Joannopoulos JD. Direct calculation of thermal emission for three-dimensionally periodic photonic crystal slabs. *Physical Review E* 2006; **74**: 1–9.

5 Green MA. Time-Asymmetric Photovoltaics. *Nano Letters* 2012; **12**: 5985–5988.

6 Park Y, Zhao B, Fan S. Reaching the Ultimate Efficiency of Solar Energy Harvesting with a Nonreciprocal Multijunction Solar Cell. *Nano Letters* 2022; **22**: 448–452.

7 Lorentz HA. The theorem of Poynting concerning the energy in the electromagnetic field and two general propositions concerning the propagation of light. *Amsterdammer Akademie der Wetenschappen* 1896; **4**: 176.

8 Shayegan KJ, Biswas S, Zhao B, Fan S, Atwater HA. Direct observation of the violation of Kirchhoff's law of thermal radiation. *Nature Photonics* 2023; **17**: 891–896.

9 Remer L, Mohler E, Grill W, Lüthi B. Nonreciprocity in the optical reflection of magnetoplasmas. *Physical Review B* 1984; **30**: 3277–3282.

10 Snyder WC, Wan Z, Li X. Thermodynamic constraints on reflectance reciprocity and Kirchhoff's law. *Applied Optics* 1998; **37**: 3464.

11 Zhu L, Fan S. Near-complete violation of detailed balance in thermal radiation. *Physical Review B* 2014; **90**: 220301.

12 Shayegan KJ, Zhao B, Kim Y, Fan S, Atwater HA. Nonreciprocal infrared absorption via resonant magneto-optical coupling to InAs. *Science Advances* 2022; **8**: eabm4308.

13 Liu M, Xia S, Wan W, Qin J, Li H, Zhao C *et al.* Broadband mid-infrared non-reciprocal absorption using magnetized gradient epsilon-near-zero thin films. *Nature Materials* 2023; **22**: 1196–1202.

14 Shayegan KJ, Hwang JS, Zhao B, Raman AP, Atwater HA. Broadband nonreciprocal thermal emissivity and absorptivity. *Light: Science & Applications* 2024; **13**: 176.

15 Pajovic S, Tsurimaki Y, Qian X, Chen G, Boriskina SV. Nonreciprocal reflection of mid-infrared light by highly doped InAs at low magnetic fields. *Optics Express* 2025; **33**: 8661–8674.

16 Nabavi B, Jafari Ghalekohneh S, Shayegan KJ, Tervo EJ, Atwater H, Zhao B. High-Temperature Strong Nonreciprocal Thermal Radiation from Semiconductors. *ACS Photonics* 2025; **12**: 2767–2774.

17 Zhang Z, Kalantari Dehaghi A, Ghosh P, Zhu L. Observation of Strong Nonreciprocal Thermal Emission. *Physical Review Letters* 2025; **135**: 16901.

18 Miyazaki HT, Ikeda K, Kasaya T, Yamamoto K, Inoue Y, Fujimura K *et al.* Thermal emission of two-color polarized infrared waves from integrated plasmon cavities. *Applied Physics Letters* 2008; **92**: 141114.

19 Starko-Bowes R, Dai J, Newman W, Molesky S, Qi L, Satija A *et al.* Dual-band quasi-coherent radiative thermal source. *Journal of Quantitative Spectroscopy and Radiative Transfer* 2018; **216**: 99–104.

20 Wang X, Sentz T, Bharadwaj S, Ray SK, Wang Y, Jiao D *et al.* Observation of nonvanishing optical helicity in thermal radiation from symmetry-broken metasurfaces. *Sci Adv* 2023; **9**: eade4203.

21 Nguyen A, Hugonin J-P, Coutrot A-L, Garcia-Caurel E, Vest B, Greffet J-J. Large circular dichroism in the emission from an incandescent metasurface. *Optica* 2023; **10**: 232.

22 Guo T, Jiang C, Zhang X, Xu G, Zarei MR, Wang J *et al.* Nanophotonics Engineering Polarization Manipulation in Thermal Radiations: A Review. *Advanced Optical Materials* 2026; **14**: e03724.

23 Fang J, Wang M, Liu T, Yue J, Sun X, Wu Y *et al.* Dual-polarization strong nonreciprocal thermal radiation with silicon-based nanopore arrays. *International Journal of Thermal Sciences* 2024; **195**: 108602.

24 Zou H, Wang B, Wu J. Polarization-independent nonreciprocal thermal radiation by cylindrical grating structure. *International Journal of Heat and Mass Transfer* 2024; **231**: 125819.

25 Huang X, Wang B, Zhou J, Ye J. Silicon cylinders hollowed for nonreciprocal thermal radiation under a magnetic field of 1 T. *International Journal of Heat and Mass Transfer* 2025; **241**: 126738.

26 Liu R, Wang B, Ye J, Wu X, Zhang F, Wang Q *et al.* Circular hole Si based on graphene/InAs for nonreciprocal thermal radiation at angle of 13° for mid-infrared region. *Diamond and Related Materials* 2025; **159**: 112839.

27 Xia S, Liu M, Wan W, Wang J, Yang W, Wang C *et al.* Dual-polarization control of broadband nonreciprocal thermal radiation by combining local and nonlocal metasurfaces. 2026. doi:10.48550/arXiv.2608.05640.

28 Johnson SG, Fan S, Villeneuve PR, Joannopoulos JD, Kolodziejski LA. Guided modes in photonic crystal slabs. *Phys Rev B* 1999; **60**: 5751–5758.

29 Seeger K. *Semiconductor Physics: An Introduction*. Springer-Verlag Berlin Heidelberg: New York, 2004.

30 Johnson S, Joannopoulos J. Block-iterative frequency-domain methods for Maxwell's equations in a planewave basis. *Opt Express* 2001; **8**: 173.

31 Mandelshtam VA, Taylor HS. Harmonic inversion of time signals and its applications. *The Journal of Chemical Physics* 1997; **107**: 6756–6769.

32 Johnson SG. Harminv. 2004.https://github.com/NanoComp/harminv.

33 Oskooi AF, Roundy D, Ibanescu M, Bermel P, Joannopoulos JD, Johnson SG. Meep: A flexible free-software package for electromagnetic simulations by the FDTD method. *Computer Physics Communications* 2010; **181**: 687–702.

34 Onsager L. Reciprocal Relations in Irreversible Processes. I. *Physical Review* 1931; **37**: 405–426.

35 Casimir HBG. On Onsager's principle of microscopic reversibility. *Reviews of Modern Physics* 1945; **17**: 343–350.

36 Camley RE. Nonreciprocal surface waves. *Surface Science Reports* 1987; **7**: 103–187.

37 Zhao B, Shi Y, Wang J, Zhao Z, Zhao N, Fan S. Near-complete violation of Kirchhoff's law of thermal radiation with a 0.3 T magnetic field. *Optics Letters* 2019; **44**: 4203.

38 Franta D, Nečas D, Zajíčková L, Ohlídal I, Stuchlík J. Advanced modeling for optical characterization of amorphous hydrogenated silicon films. *Thin Solid Films* 2013; **541**: 12–16.

39 Chang H-J, Kim S-H, Lee Y-H, Kartalov EP, Scherer A. A photonic-crystal optical antenna for extremely large local-field enhancement. *Opt Express* 2010; **18**: 24163.

40 Levinshtein M, Rumyantsev S, Shur M. *Handbook Series on Semiconductor Parameters: Volume 1: Si, Ge, C (Diamond), GaAs, GaP, GaSb, InAs, InP, InSb*. WORLD SCIENTIFIC, 1996 doi:10.1142/2046-vol1.

41 Celanovic I, Jovanovic N, Kassakian J. Two-dimensional tungsten photonic crystals as selective thermal emitters. *Applied Physics Letters* 2008; **92**: 193101.

42 Ilic O, Bermel P, Chen G, Joannopoulos JD, Celanovic I, Soljačić M. Tailoring high-temperature radiation and the resurrection of the incandescent source. *Nature Nanotechnology* 2016; **11**: 320–324.

43 Lin SY, Fleming JG, Hetherington DL, Smith BK, Biswas R, Ho KM *et al.* A three-dimensional photonic crystal operating at infrared wavelengths. *Nature* 1998; **394**: 251–253.

44 Fleming JG, Lin SY, El-Kady I, Biswas R, Ho KM. All-metallic three-dimensional photonic crystals with a large infrared bandgap. *Nature* 2002; **417**: 52–55.

45 Lin SY, Fleming J, Chow E, Bur J, Choi K, Goldberg A. Enhancement and suppression of thermal emission by a three-dimensional photonic crystal. *Physical Review B* 2000; **62**: R2243–R2246.

46 García-Santamaría F, Xu M, Lousse V, Fan S, Braun PV, Lewis JA. A Germanium Inverse Woodpile Structure with a Large Photonic Band Gap. *Advanced Materials* 2007; **19**: 1567–1570.

47 Peng S, Zhang R, Chen VH, Khabiboulline ET, Braun P, Atwater HA. Three-Dimensional Single Gyroid Photonic Crystals with a Mid-Infrared Bandgap. *ACS Photonics* 2016; **3**: 1131–1137.

48 Greffet J-J, Carminati R, Joulain K, Mulet J-P, Mainguy SS, Chen Y. Coherent emission of light by thermal sources. *Nature* 2002; **416**: 61–64.

49 Yao Y, Ikeda N, Kuroda T, Mano T, Koyama H, Sugimoto Y *et al.* Mid-IR Dirac-cone dispersion relation materialized in SOI photonic crystal slabs. *Opt Express* 2020; **28**: 4194.

50 Vaidya S, Noh J, Cerjan A, Jörg C, Von Freymann G, Rechtsman MC. Observation of a Charge-2 Photonic Weyl Point in the Infrared. *Phys Rev Lett* 2020; **125**: 253902.

51 Yi F, Liu MQ, Wang NN, Wang BX, Zhao CY. Near-field observation of mid-infrared edge modes in topological photonic crystals. *Applied Physics Letters* 2023; **123**: 081110.

52 Franta D, Nečas D, Ohlídal I, Giglia A. Dispersion model for optical thin films applicable in wide spectral range. In: Duparré A, Geyl R (eds). *Proceedings of SPIE*. Jena, Germany, 2015, p 96281U.

53 Adambukulam C, Sewani VK, Stemp HG, Asaad S, Mądzik MT, Morello A *et al.* An ultra-stable 1.5 T permanent magnet assembly for qubit experiments at cryogenic temperatures. *Review of Scientific Instruments* 2021; **92**: 085106.

54 Chubar O, Elleaume P, Chavanne J. A three-dimensional magnetostatics computer code for insertion devices. *J Synchrotron Rad* 1998; **5**: 481–484.

55 Hall C, Abell D, Banerjee A, Chubar O, Edelen J, Keilman M *et al.* Recent developments to the Radia magnetostatics code for improved performance and interface. *J Phys: Conf Ser* 2022; **2380**: 012025.

56 Truong R. *Permanent magnet assembly for ion trapping*. 2025.

# Supplementary Information for

# "Dual-polarized, mid-infrared nonreciprocal absorption"

Simo Pajovic[1,2]*, Yiting Zhao[1], Yae-Chan Lim[1], Ruzan Sokhoyan[1], Harry A. Atwater[1]†

[1] *Department of Applied Physics and Materials Science, California Institute of Technology, United States of America*

[2] *Resnick Sustainability Institute, California Institute of Technology, United States of America*

*pajovics@caltech.edu

†haa@caltech.edu

**Table of Contents**

## I. Fabrication

Figure S1 illustrates the fabrication process of the PhC-on-InAs design. Details of the fabrication were described in **Methods** in the main text. Optical microscope images following critical steps in the fabrication process are shown in Fig. S2.

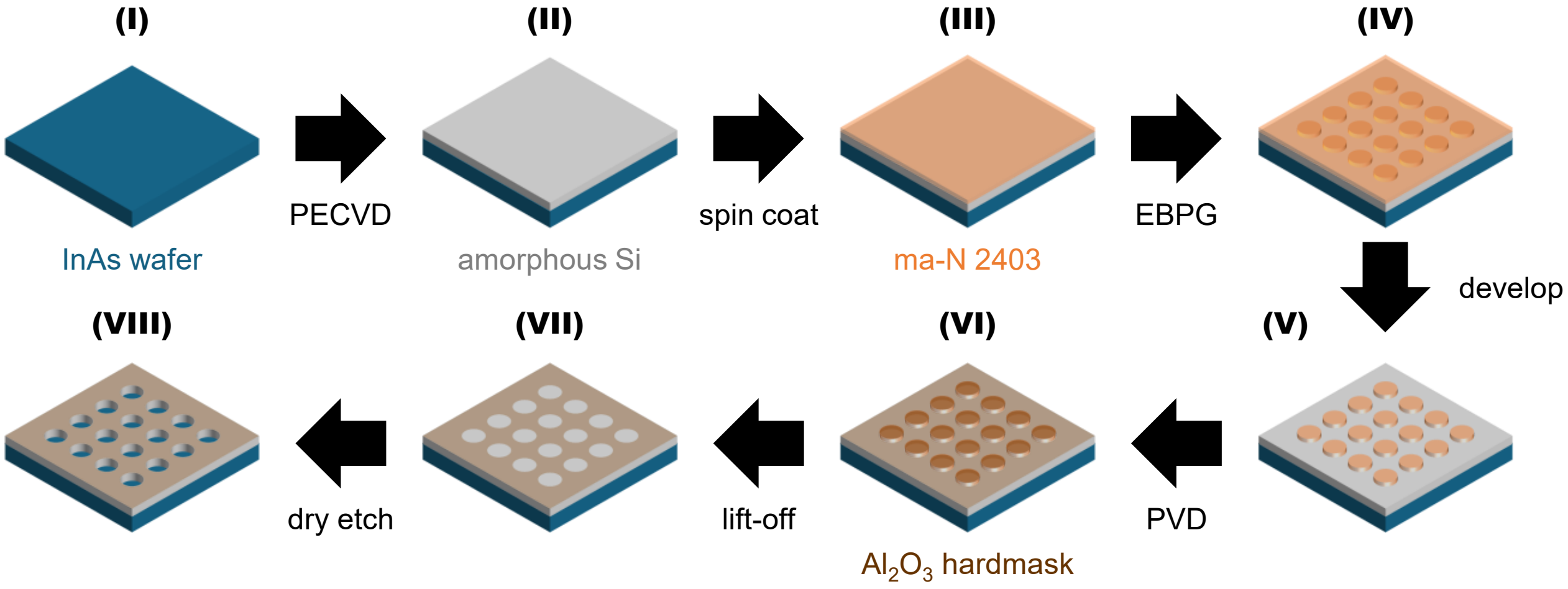

**Figure S1: Fabrication of the PhC-slab-on-InAs design.** (Clockwise) (I) Start with a pristine InAs wafer. (II) Deposit a-Si using plasma-enhanced chemical vapor deposition (PECVD). (III) Spin coat ma-N 2403 electron beam resist. (IV) Pattern resist using electron beam pattern generator (EBPG). (V) Develop resist in MF-319. (VI) Deposit hardmask using physical vapor deposition (PVD, or electron-beam evaporator). (VII) Lift-off in Remover PG, exposing holes in the hardmask. (VIII) Dry etch a-Si using the Bosch process, resulting in an a-Si PhC slab on top of the InAs wafer.

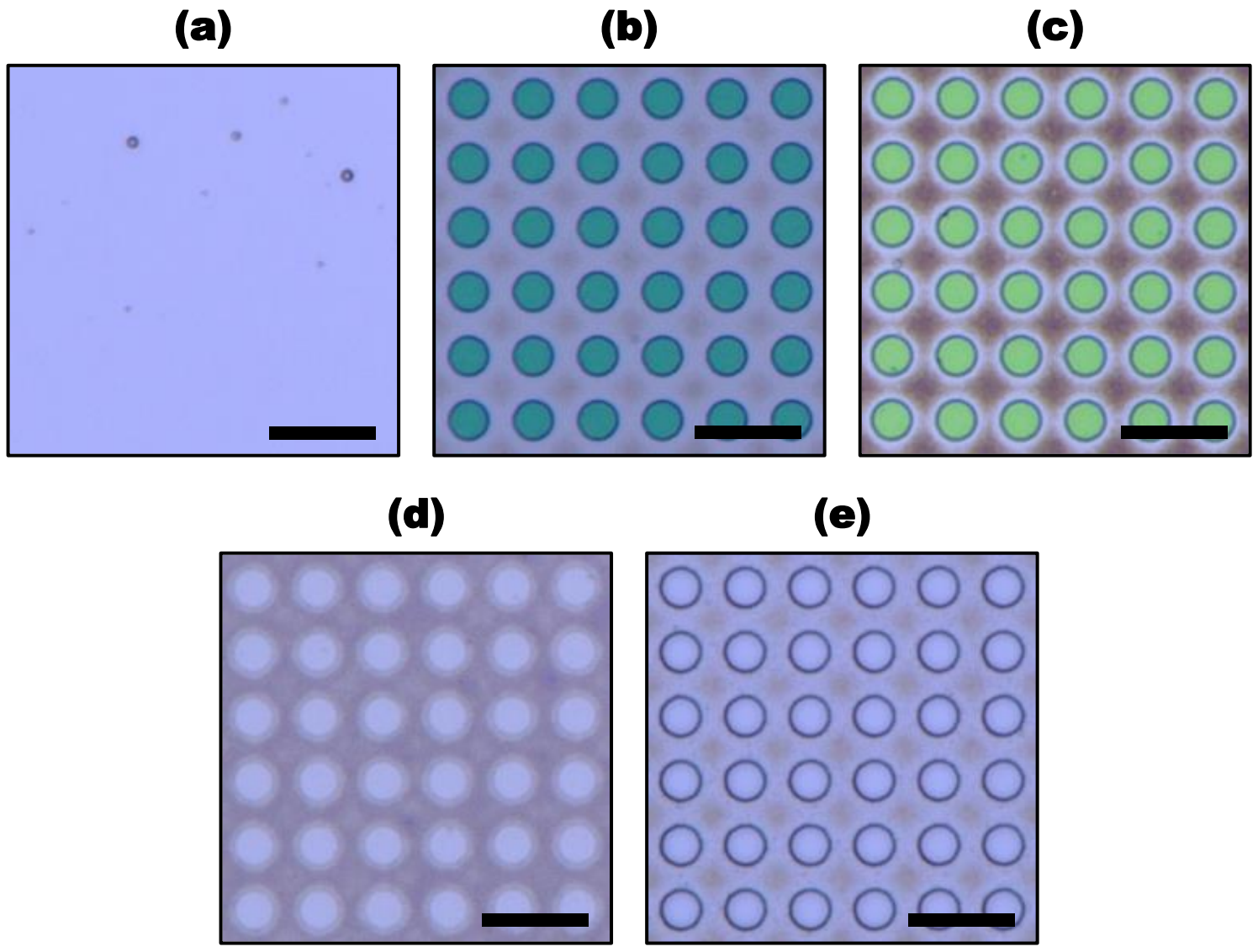

**Figure S2: Optical microscope images after different steps in the fabrication process.** Original microscope images cropped to show a few unit cells. (a) a-Si film after deposition, showing relatively small bubbles in the film (after step II in Fig. S1). (b) Developed ma-N 2403 resist (step V). (c) Hardmask before lift-off (after step VI). (d) After lift-off (step VII). (e) After dry etching (step VIII). All scale bars (black) are 10 μm.

## II. Ellipsometry

We used ellipsometry (IR-VASE Mark II, J.A. Woollam) to measure the zero-field dielectric functions of the InAs wafer and a-Si layers (where it is assumed that a-Si does not support magneto-optical effects anyway). The results are shown in Figs. S3 and S4, including the measured (markers) and fitted (lines) $\Delta$ and $\Psi$, as well as the fitted dielectric functions following Eqs. (8) and (9) in the main text.

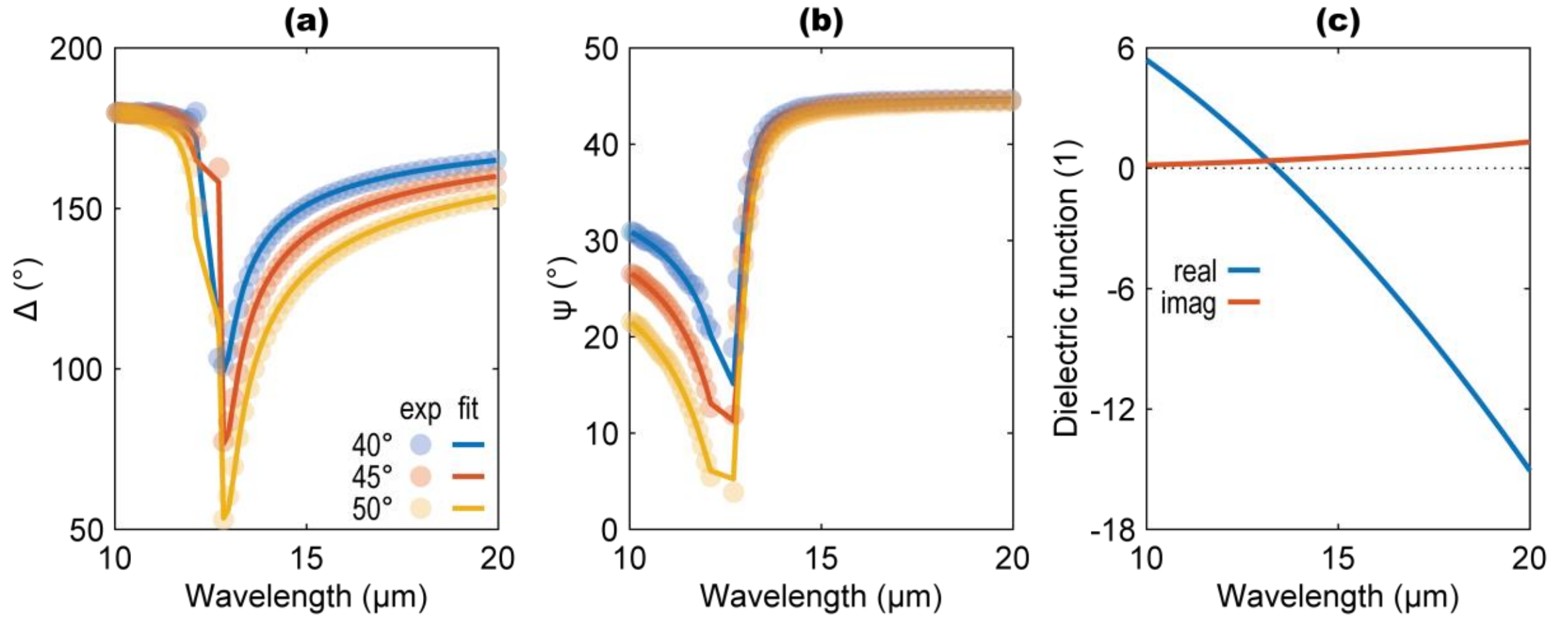

**Figure S3: Zero-field dielectric function of InAs.** Ellipsometry (a) $\Delta$ and (b) $\Psi$ of the InAs wafer used in this work as a function of angle of incidence, with zero applied magnetic field. Measured data are indicated by the markers, while the data fit is indicated by the lines. Different colors indicate the angle of incidence per the legend. (c) Real (blue) and imaginary (orange) dielectric function of InAs fitted using Eq. (8) in the main text. The ENZ wavelength is approximately 13.2 μm.

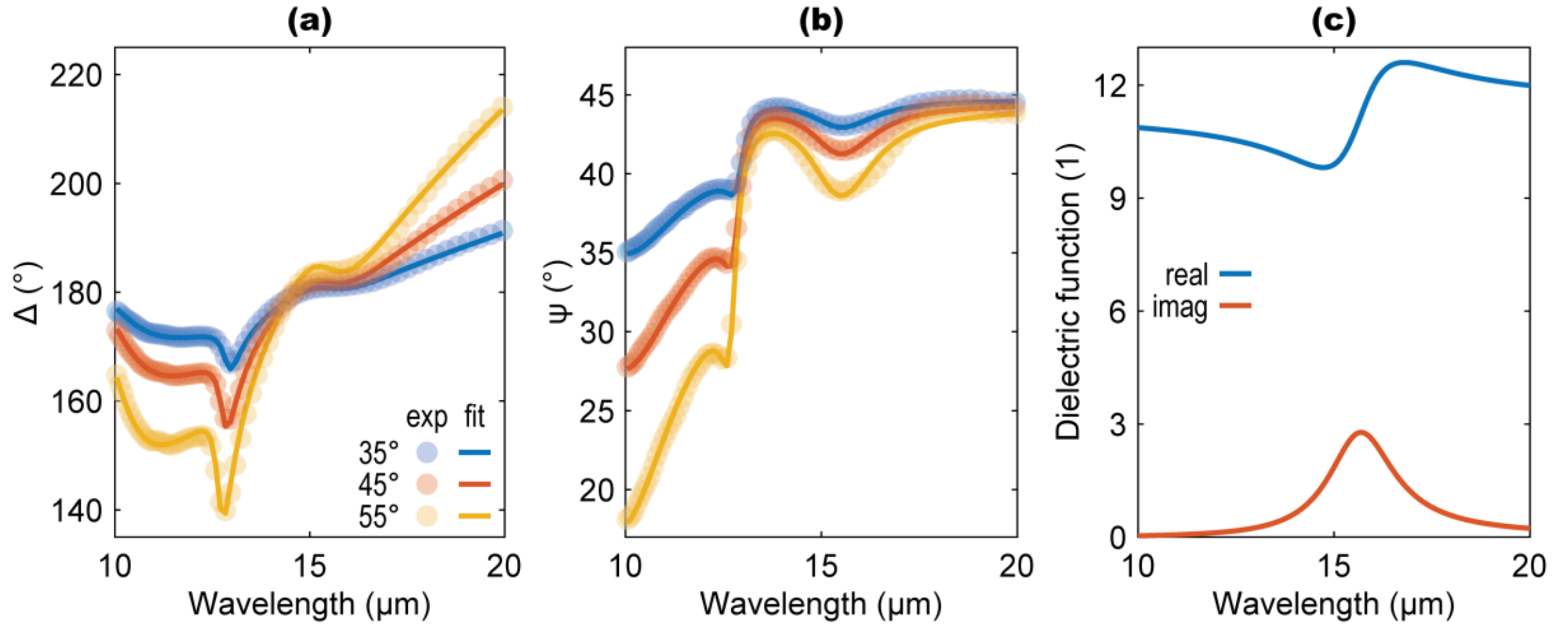

**Figure S4: Dielectric function of a-Si.** (a) $\Delta$ and (b) $\Psi$ of the same InAs wafer as in Fig. S3 with approximately 1.477 μm of a-Si deposited on top. (c) Real and imaginary dielectric function of a-Si fitted using Eq. (9) in the main text.

## III. Magnetic field characterization

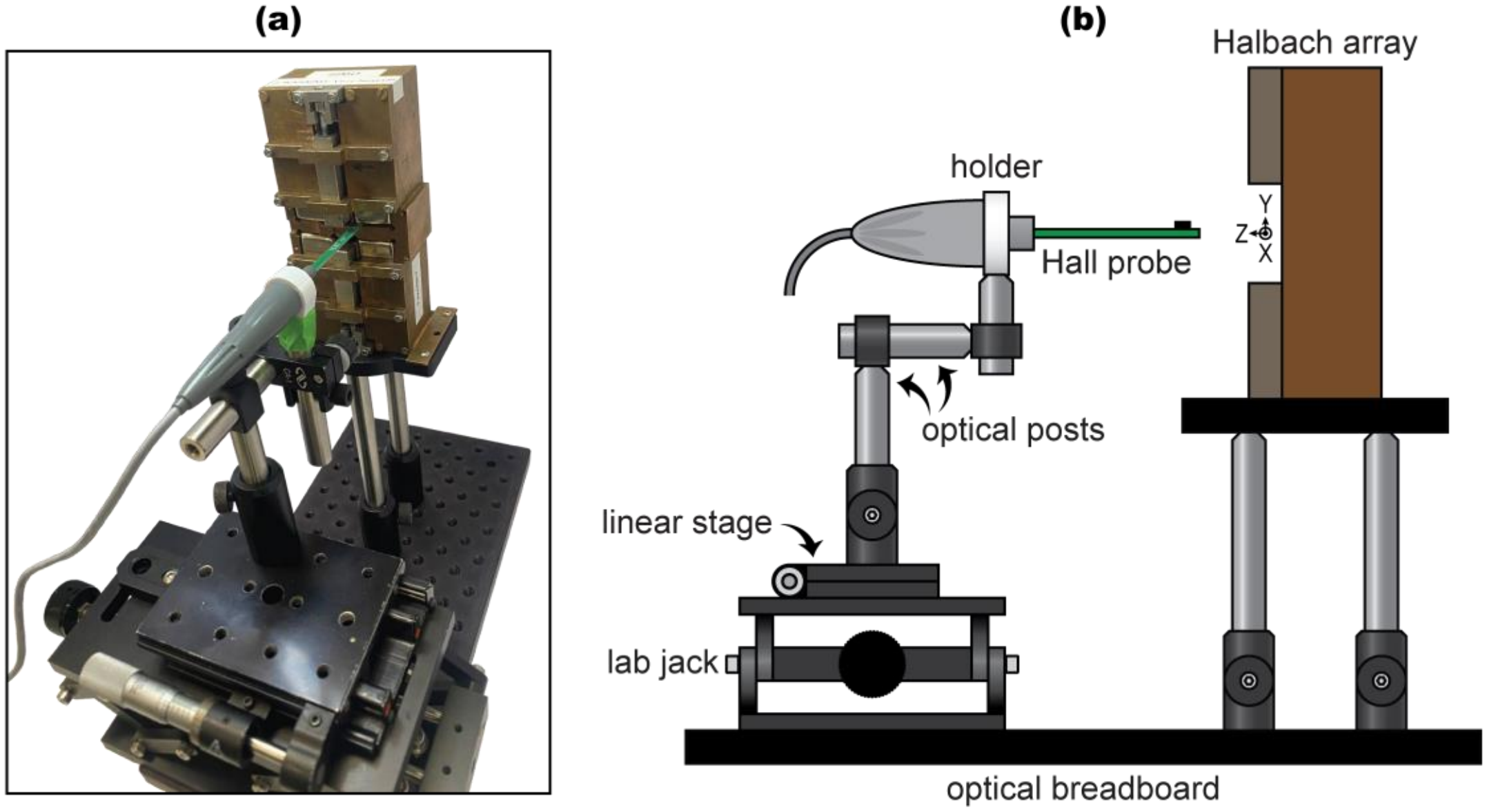

**Figure S5: Magnetic field measurement.** (a) Photograph of the experimental setup. (b) Schematic. The setup consisted of a lab jack for $y$ translation and a linear stage for $x$ translation. The initial position of the Hall probe could be adjusted using a set of sliding optical posts, including a custom 3D printed holder to interface with the optical posts. The coordinate axes are shown in the middle of the gap between the pole pieces of the Halbach array.

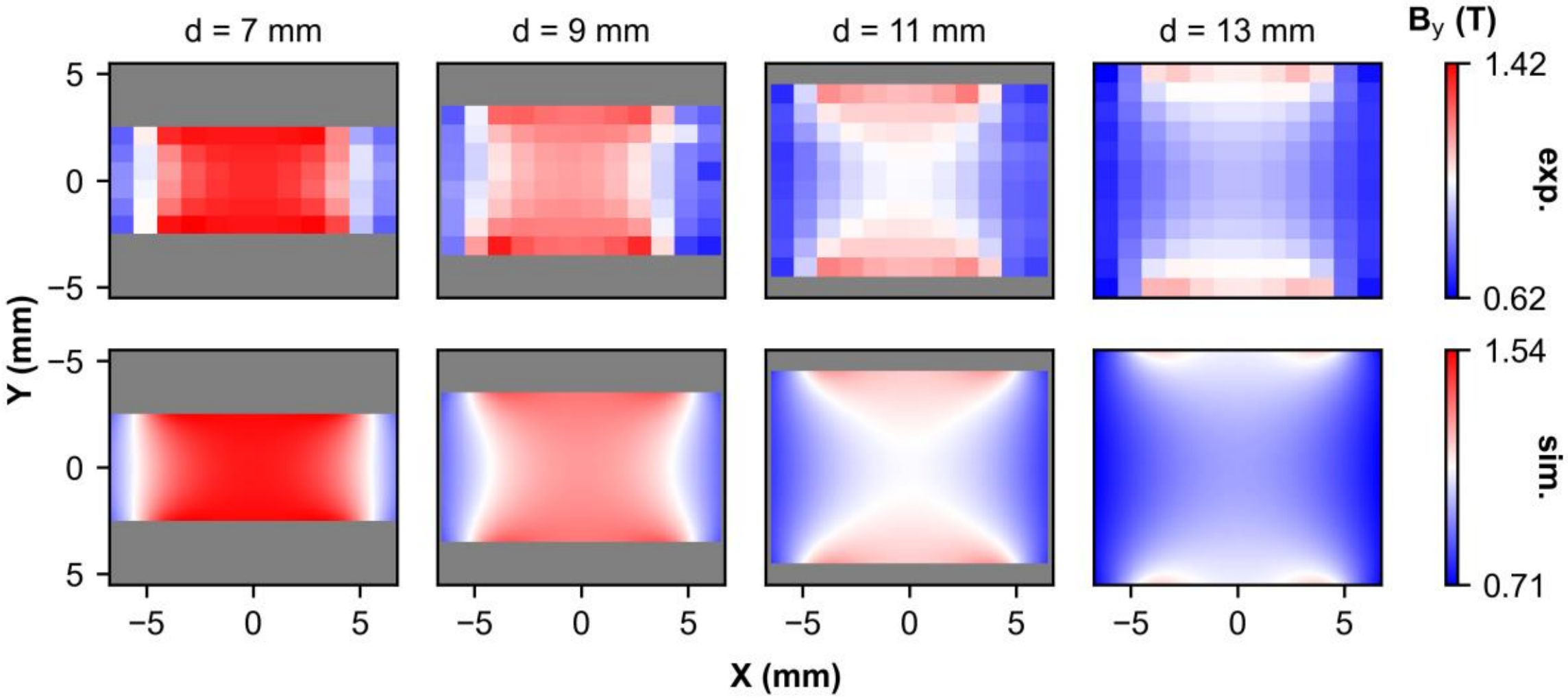

**Figure S6: Experimental and simulated magnetic field distributions between the Supermendur pole pieces.** The top row shows the experimental magnetic field distributions measured using the setup shown in Fig. S5, while the bottom row shows the simulated magnetic field distributions calculated using Radia[1,2]. Across rows, the gap width between the pole pieces, $d$, increases from left to right. Overall, these results show good agreement with each other, and discrepancies in magnitude can be explained by slight inaccuracies in the simulated geometry and differences between the simulated and experimental remanent magnetizations of the permanent magnets[3].

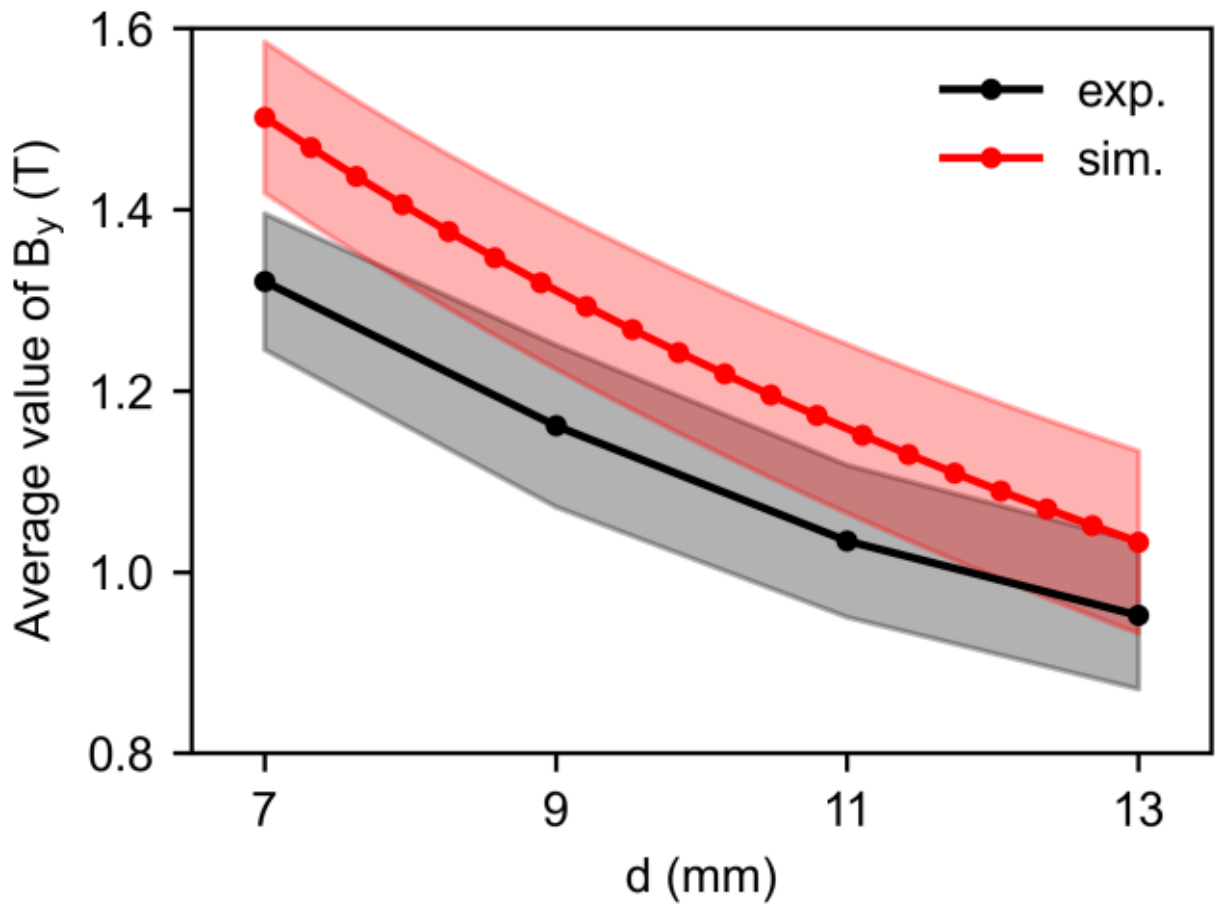

**Figure S7: Experimental and simulated average magnetic field.** The average values of the magnetic field distributions shown in Fig. S6, calculated as a function of the gap width $d$. The black curve is experimental, and the red curve is simulated. The average is taken over an area in the $xy$-plane slightly smaller than the gap itself (padded by 0.25 mm). The shaded areas are the standard deviation, showing how much the magnetic field varies within the gap. The beam spot size does not fill up the entire gap, so this standard deviation overestimates the nonuniformity of the magnetic field.

## IV. Discussion on fabrication and experimental imperfections

In this section, we discuss possible sources of fabrication and experimental imperfections that may have led to the discrepancies between theory and experiments observed in the main text. Supporting FEM simulations are shown and discussed in Sec. V of this document.

*Background absorptance contrast.* As discussed in the main text, we observe a broadband, nonresonant, nonzero absorptance contrast in the 14–18 μm spectral range, which we attribute to unintended scattering or diffraction originating from fabrication imperfections. The sample has defects due to imperfect lift-off, which can be local (at the level of a single unit cell, such as a pillar of developed resist that failed to lift off) or macroscopic (delamination of the hardmask over a few tens of unit cells). These macroscopic defects can be especially detrimental, as they are much longer than the wavelength of light and make the optical response of the sample nonuniform.

*Blueshifting.* This blueshifting could have been caused, for example, by etching-related defects such as sidewall sloping or roughness. Based on SEM images of the sample, the sidewalls of the holes appeared to be straight, and sidewall roughness, while present, appeared to be deeply subwavelength (i.e., $\ll 10$ μm). Still, using FEM simulations, we found that modest sidewall sloping could have slightly shifted the peaks (Supplementary Fig. S9).

*Differences in absorptance magnitude.* Although the magnitude of the experimental absorptance contrast agrees well with theory, the magnitude of the absorptance itself is slightly different. We attribute this to scattering in the vicinity of the sample (e.g., due to the pole pieces) and background signal from the ambient environment. This is supported by the fact that we observed high sensitivity of the background absorptance contrast to small variations in the experimental setup. Slight apodization of the beam, the precise location of the beam spot (linked to the sample's macroscopic defects), and angular drift due to the heavy Halbach array were all factors that could have affected the absorptance. From run to run, these were normalized using a gold mirror cut to the same size as the sample as a reference, but ultimately, FEM simulations are limited in their ability to mimic every detail of the experiments.

*Additional imperfections.* We also studied whether magnetic field variations and azimuthal angle offset could have impacted our results. Our FEM simulations assumed the applied magnetic field was uniform via the dielectric tensor of InAs, whereas in experiments, it was spatially dependent (weakly so over the beam spot size, see Fig. 4(a)(i)). Nevertheless, we found that varying $B$ within approximately 0.2 T of the spatially-averaged value only strengthened or weakened the nonreciprocal effect without shifting the peaks and valleys (Supplementary Fig. S10). We also considered whether the sample was misaligned in the azimuthal direction due to a slight, unintentional twist about the $z$-axis, but we also found via FEM simulations that this is insignificant up to a few degrees of azimuthal angle offset. Even then, this offset would have been noticeable in experiments via a nonzero cross-polarized reflectance, $R_{sp}$ and $R_{ps}$, but these were close to zero with no clear trend, whereas theory predicts that they should be nonzero near the resonances of the sample (Supplementary Fig. S11).

## V. Simulated effects of experimental imperfections on results

In this section, we simulate the effects of different experimental imperfections on dual-polarized nonreciprocal absorptance. In all the following calculations, we assume the same geometry and material properties as the sample in the main text., Compared to the main text, we truncate our analysis to the 11–14 μm spectral range, which we identify as the approximate ENZ region (with the ENZ wavelength being 13.2 μm). This is because nonreciprocity is strongest in this region, allowing us to expedite the simulations while still understanding how experimental imperfects may have affected the original intended design.

*Hardmask*. The $Al_2O_3$ hardmask on top of the as-fabricated structure was approximately 50 nm thick. For this reason, we assumed that it was optically thin and only had a negligible effect on the optical response of the sample, even though $Al_2O_3$ has a known absorption peak at approximately 14 μm, which in turn makes it lossy over the spectral range of interest. This assumption also facilitated FEM simulations using COMSOL Multiphysics, as we found that meshing such a fine feature significantly increased computational cost. To check this assumption, we calculated the absorptance and absorptance contrast of the as-fabricated structure with and without the hardmask, using optical properties from the literature for $Al_2O_3$[4]. The results are shown in Fig. S8, with the markers indicating the calculations from the main text without the hardmask and the lines indicating additional calculations with the hardmask. As expected, the effect on the optical response is minimal, explained by the fact that $Al_2O_3$ itself is not gyrotropic at room temperature in the mid-infrared and that extra absorption from the lossy thin film should have a similar effect in both light propagation directions. Confirming the negligible effects of the hardmask greatly simplified simulations.

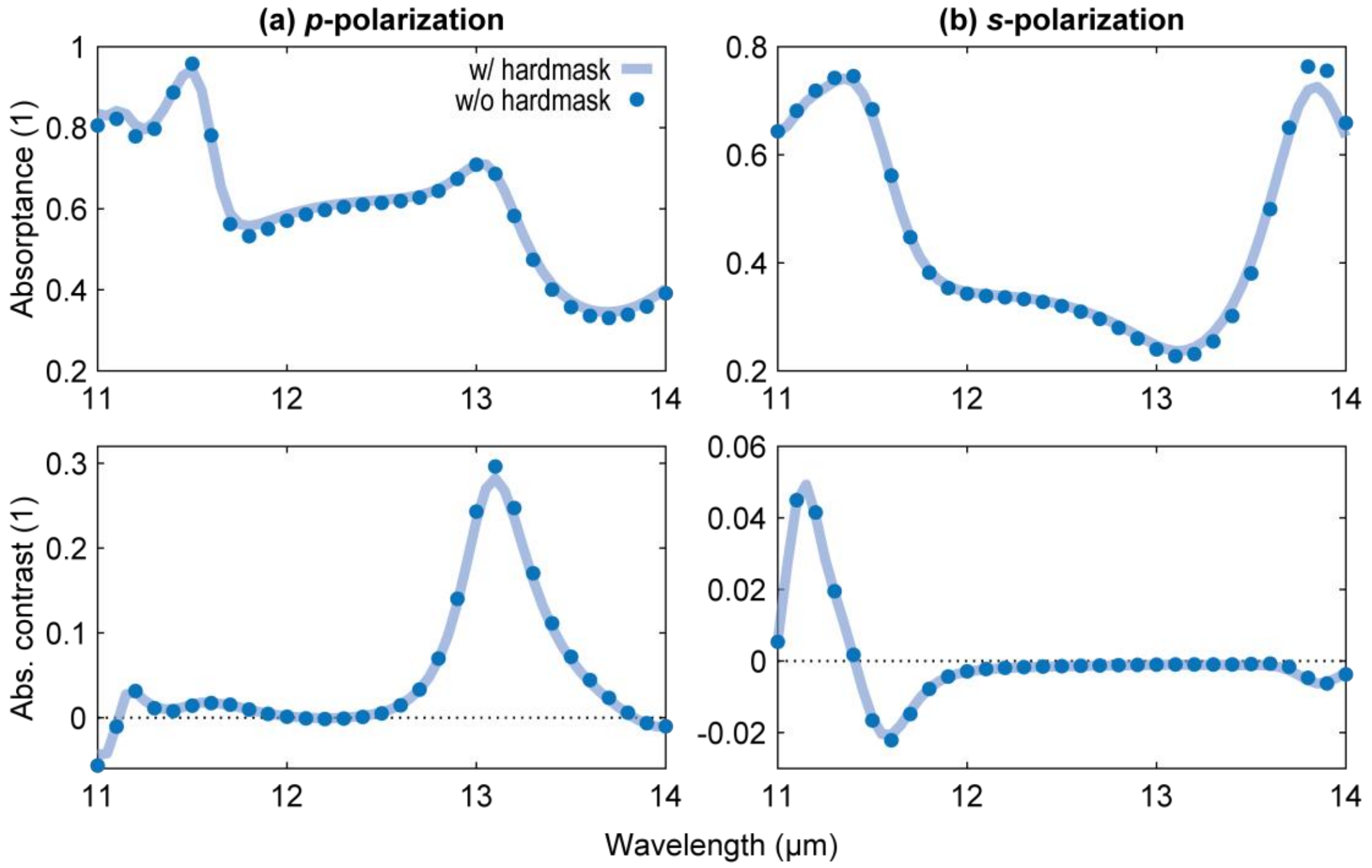

**Figure S8: Effect of $Al_2O_3$ hardmask.** Simulated absorptance $a_\ell(\lambda, \theta, B)$ and absorptance contrast $\Delta a_\ell(\lambda, \theta, B) = a_\ell(\lambda, -\theta, B) - a_\ell(\lambda, \theta, B)$ (Eq. (6) in the main text), plotted as a function of wavelength $\lambda$ and polarization $\ell$, with (lines) and without (markers) the $Al_2O_3$ hardmask. $\theta = 50°$ and $B = 1$ T. The left column (a) shows the case of $p$-polarized light, while the right (b) shows $s$-polarized. The hardmask is assumed to be 50 nm thick.

*Sidewall sloping*. Although we did not see any clearly visible sidewall sloping in SEM and optical microscope images of our structure, we nevertheless considered it possible that it could have affected the optical response of the sample. We simulated a range of sidewall sloping angles, from a gentle, almost imperceptible slope (0.5°) to a more aggressive one (10°). In constructing the geometry, we assumed the diameter at the bottom of the hole was that of the as-fabricated structure (4.053 μm) and the walls sloped outward with angle $\alpha$. The results are shown in Fig. S9. Compared to the baseline of straight sidewalls (blue markers), sidewall sloping seems to significantly affect the optical response when the angle is quite steep ($\alpha \geq 5°$). We note that increasing $\alpha$ tends to blueshift the resonant peaks, and we observed that our experimentally observed resonant peaks were at shorter wavelengths compared to our simulations with vertical sidewalls. Therefore, it is plausible that sidewall sloping of a few degrees was present but difficult to assess with the images we had.

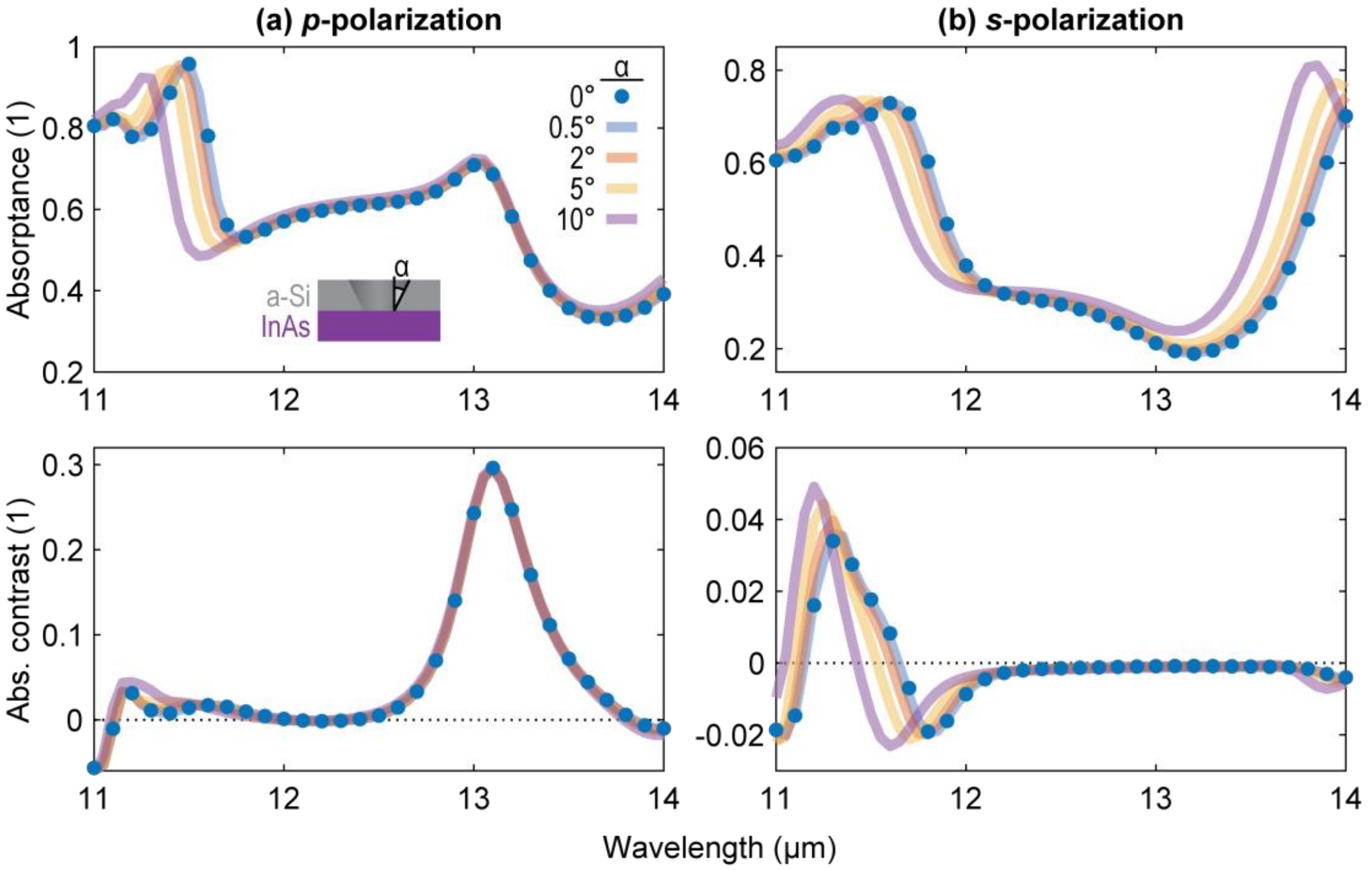

**Figure S9: Effect of sidewall sloping.** Simulated $a_\ell(\lambda, \theta, B)$ and $\Delta a_\ell(\lambda, \theta, B)$, plotted as a function of wavelength $\lambda$, polarization $\ell$, and sidewall sloping angle $\alpha$. (defined in the inset drawing). $\theta = 35°$ and $B = 1$ T. Left column (a) shows $p$-polarization, while right column (b) shows $s$-polarization. Markers indicate the case of $\alpha = 0°$ or vertical sidewalls (as in the main text), while different-colored lines indicate $\alpha \neq 0°$.

*Magnetic field variations.* In our simulations, we assume a uniform magnetic field with strength equal to the average value from experiments, plotted in Fig. S7. In practice, however, the magnetic field has some spatial dependence within the gap, as shown in Fig. S6. To simulate our sample over such a large area accurately accounting for the magnetic field distribution would be extremely computationally expensive, given that our unit cell has a period of 6.105 μm and the length scale of these spatial variations is on the order of mm (nearly 3 orders of magnitude difference). Nevertheless, to probe the effect of magnetic field variations, we simulated our structure for different values of the uniform magnetic field within approximately two standard deviations of the averages shown in Fig. S7, or double the approximately 0.1 T wide error bars. The results are shown in Fig. S10. As expected, changing the magnetic field by up to 0.2 T, which is significant, changes the strength of nonreciprocal absorption via the absorptance contrast without spectrally shifting the resonant peaks. This is why the lines in Fig. S10, corresponding to $B \neq 1$ T, are symmetric about the markers, corresponding to $B = 1$ T, with lower values lying below and higher values lying above.

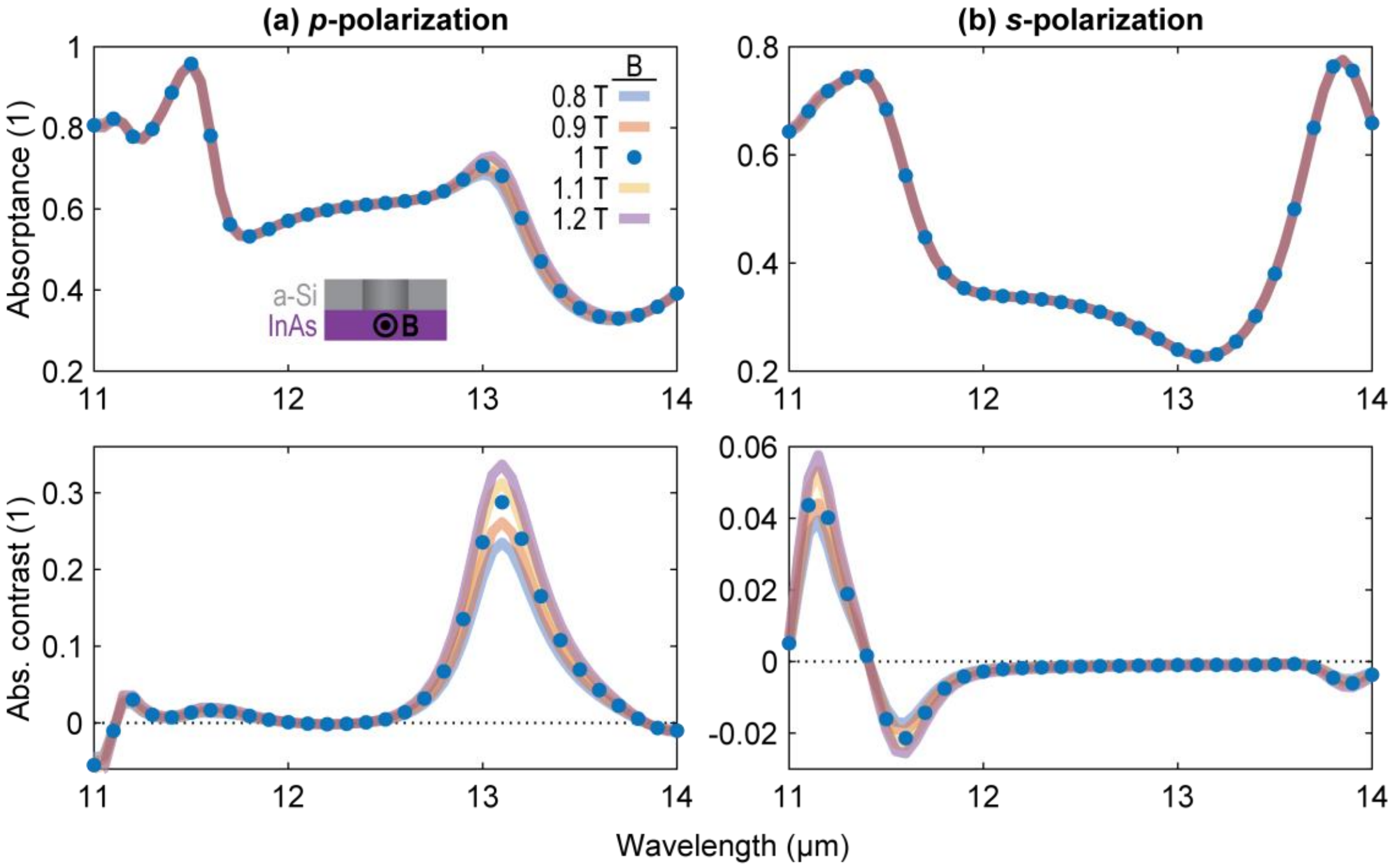

**Figure S10: Effect of magnetic field variations.** Simulated $a_\ell(\lambda, \theta, B)$ and $\Delta a_\ell(\lambda, \theta, B)$, plotted as a function of wavelength $\lambda$, polarization $\ell$, and magnetic field $B$. $\theta = 50°$. Left column (a) shows $p$-polarization, while right column (b) shows $s$-polarization. Markers indicate the case of $B = 1$ T (as in the main text), while different-colored lines indicate $B \neq 1$ T.

*Azimuthal angle offset*. Finally, we considered the possibility that the sample had a slight azimuthal angle offset during experiments, meaning the sample would have been off the Voigt configuration. To simulate this effect in COMSOL Multiphysics, we rotated both the plane of incidence and the magnetic field, as shown in the inset drawing in Fig. S11. We transform the incident wavevector, incident polarization, and dielectric tensor of InAs by the following rotation matrix:

$$\bar{\bar{R}} = \begin{bmatrix} \cos\varphi & -\sin\varphi & 0 \\ \sin\varphi & \cos\varphi & 0 \\ 0 & 0 & 1 \end{bmatrix}, \tag{S1}$$

where $\varphi$ is the azimuthal angle of incidence. This represents a rotation about the $z$-axis (consistent with the coordinate axes throughout the figures).

The results of our calculations are shown in Fig. S11. Importantly, when $\varphi \neq 0°$, there is polarization conversion upon reflection, meaning, e.g., an incoming $p$-polarized wave scatters into outgoing $p$- and $s$-polarized waves, resulting in two reflectances, $R_{\ell\ell}$ and $R_{\ell m}$, where $\ell, m = s, p$. For this reason, we plot reflectance instead of absorptance in Fig. S11, although we plot $\Delta a_{\ell\ell}(\lambda, \theta, B) = R_{\ell\ell}(\lambda, \theta, B) - R_{\ell\ell}(\lambda, \theta, B)$ for the absorptance contrast. The effect of azimuthal angle offset becomes significant at $\varphi \geq 5°$, with performance degradation originating from scattering into the additional polarization channel that cannot be coupled into in the Voigt configuration. We note that $R_{\ell m}(\lambda, \theta, B)$ is nonzero in the vicinity of the absorption peaks (or reflection valleys), providing a clear signature of azimuthal angle offset to look out for in experiments. However, we did not observe this signature and found that $R_{\ell m}(\lambda, \theta, B) \approx 0$. It is plausible that a small azimuthal angle offset was present (say, $\varphi \lesssim 2°$ based on the data in Fig.

S11), but this would be extremely difficult to confirm. Either way, we believe this offset was not a significant factor in our experiments.

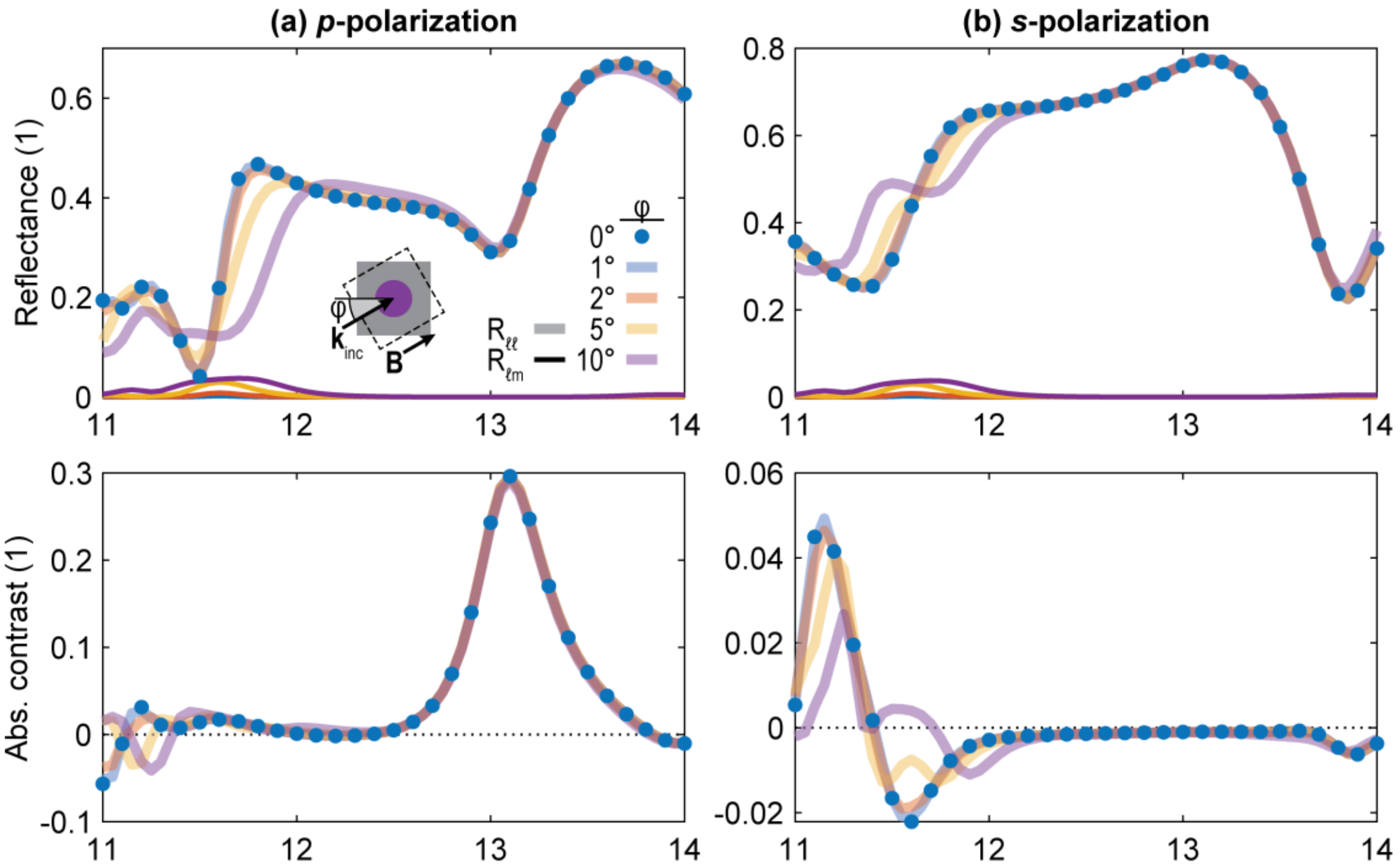

**Figure S11: Effect of azimuthal angle offset.** Simulated reflectance $R_{\ell m}(\lambda, \theta, B)$ and $\Delta a_{\ell\ell}(\lambda, \theta, B)$, plotted as a function of wavelength $\lambda$, incoming and outgoing polarizations $\ell$ and $m$, and azimuthal angle of incidence $\varphi$ (defined in the inset drawing). $\theta = 50°$ and $B = 1$ T. Left column (a) shows $p$-polarization, while right column (b) shows $s$-polarization. Markers indicate the case of $\varphi = 0°$ or the Voigt configuration (as in the main text), while different-colored lines indicate $\varphi \neq 0°$. Additionally, thicker, translucent lines correspond to $\ell = m$, while thinner, opaque lines correspond to $\ell \neq m$. Note that for the absorptance contrast, we ignore the case of $\ell \neq m$.

## VI. Dummy Halbach array

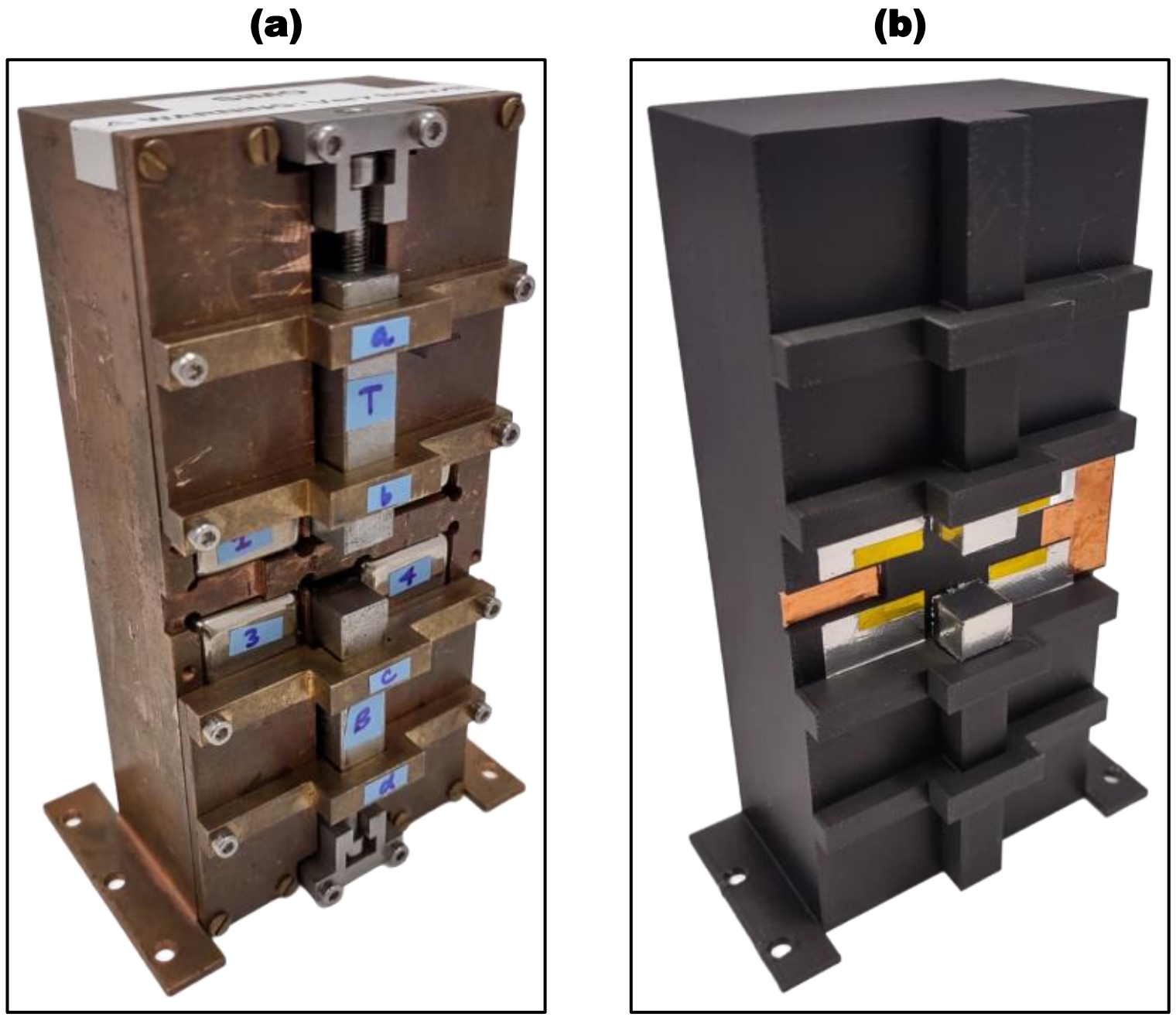

**Figure S12: Dummy Halbach array for zero-field absorptance measurements.** (a) The real Halbach array used to measure the magnetic field-dependent absorptance. (b) The dummy, used to simulate the surroundings of the real one when measuring zero-field absorptance. Aluminum foil and copper tape are used to mimic the reflective surfaces of the magnets and the copper housing of the real Halbach array.

## VII. Uncertainty propagation

To calculate the error bars shown in Fig. 4(d), we propagated the uncertainty reported by the ellipsometer, which was the standard deviation of 40 scans per spectrum. Per Eq. (10) in the main text, the reflectance of the sample was estimated as the ratio of the sample intensity divided by the gold mirror intensity. Treating these as independent measurements, the uncertainty of the reflectance or, equivalently, absorptance is

$$\sigma_R = R_\ell \sqrt{\left(\frac{\sigma_{I,s}}{I_{\text{sample}}}\right)^2 + \left(\frac{\sigma_{I,m}}{I_{\text{mirror}}}\right)^2}, \tag{S2}$$

where $\sigma_R$, $\sigma_{I,s}$, and $\sigma_{I,m}$ are the uncertainties associated with the reflectance $R_\ell$, sample intensity $I_{\text{sample}}$, and gold mirror intensity $I_{\text{mirror}}$. To calculate the uncertainty of the absorptance contrast $\sigma_{\Delta a}$, we added the respective uncertainties of the forward and backward absorptances in quadrature:

$$\sigma_{\Delta a} = \sqrt{\sigma_R(B)^2 + \sigma_R(-B)^2}, \tag{S3}$$

where the function notation indicates the magnetic field direction.

We found that increasing scans per spectrum beyond 40 did not significantly affect the magnitude of the uncertainty. Likewise, increasing measurement cycles (the number of times the polarizers complete a rotation over all four combinations of incoming and outgoing polarization combinations, *pp*, *ss*, *sp*, and *ps*) had little effect. In the end, 40 scans per spectrum and 1 measurement cycle resulted in a good enough signal without taking too much time, allowing us to perform the fine angle of incidence sweeps shown in Figs. 3(d–e) and 4(c) in the main text.

## References

1 Chubar O, Elleaume P, Chavanne J. A three-dimensional magnetostatics computer code for insertion devices. *J Synchrotron Rad* 1998; **5**: 481–484.

2 Hall C, Abell D, Banerjee A, Chubar O, Edelen J, Keilman M *et al.* Recent developments to the Radia magnetostatics code for improved performance and interface. *J Phys: Conf Ser* 2022; **2380**: 012025.

3 Truong R. *Permanent magnet assembly for ion trapping*. 2025.

4 Franta D, Nečas D, Ohlídal I, Giglia A. Dispersion model for optical thin films applicable in wide spectral range. In: Duparré A, Geyl R (eds). *Proceedings of SPIE*. Jena, Germany, 2015, p 96281U.